\documentclass[preprint,12pt,nopreprintline]{elsarticle}

\usepackage{graphicx}
\usepackage{tabularx}
\usepackage{threeparttable}
\usepackage{mathtools}
\usepackage[version=4]{mhchem}
\usepackage[dvipsnames]{xcolor}
\usepackage{multirow}
\usepackage[colorlinks=true]{hyperref}
\usepackage{natbib}
\usepackage{units}

\newcommand{\rev}[1]{#1}

\begin{document}

\begin{frontmatter}

\title{\textit{qdmag}: A Python package for simulating nonequilibrium magnetization of molecules using quantum master equations}

\author[aff1]{Shuanglong Liu}
\author[aff1]{Xiao Chen}
\author[aff1]{Andrew Cupo}
\author[aff1]{James N. Fry}
\author[aff1]{\texorpdfstring{Hai-Ping Cheng\corref{cor1}}{Hai-Ping Cheng}}

\address[aff1]{Department of Physics, Northeastern University, Boston, Massachusetts 02115, USA} 

\cortext[cor1]{ha.cheng@northeastern.edu}

\begin{abstract}
We present \textit{qdmag}, a computational utility for calculating the magnetization of magnetic molecules in a time-varying external magnetic field by solving a generalized Lindblad quantum master equation with spin-phonon coupling treated as a dissipation term. The package relies on the spin Hamiltonian formalism, which includes both the magnetic exchange interactions and the zero-field splitting (ZFS) interaction. Different types of magnetic exchange interaction and ZFS terms up to 12\textsuperscript{th} order, which are written in terms of the extended Stevens operators, are supported. \textit{qdmag} enables long-time evolution of magnetization for up to a few milliseconds, which is achieved through a staircase approximation of the magnetic field as a function of time. For high-dimensional spin Hamiltonians, a lower-dimensional effective Hamiltonian is constructed to make the calculations computationally feasible. Three case studies of one-spin, two-spin, and three-spin systems are presented to demonstrate the usage of \textit{qdmag}. Our work offers a computational tool to bridge the gap between theoretical development and numerical calculations of spin dynamics in the presence of spin-phonon coupling.
\end{abstract}


\end{frontmatter}

\section{Introduction}

Magnetic molecules include organic radicals~\cite{Shu2023Radicals}, transition metal complexes~\cite{mabbs1973magnetism,Benelli2015MagMol}, and lanthanide-based complexes~\cite{Woodruff2013-LnMM,Hu2024-LnMM,Li2024-LnMM}. Organic radicals and transition metal complexes gain their magnetism mainly from the spin of unpaired electrons. The magnetism of lanthanide-based complexes originates from both unpaired electrons and unquenched orbital angular momentum.
Magnetic molecules often show low-energy spin excitations within hundreds of wavenumbers ($\textrm{cm}^{-1}$). 
Given a total spin $S$, the $2S+1$ spin states with varying $z$-projection $m_s$ are degenerate when relativistic effects, such as spin-orbit coupling, are not considered. 
Inclusion of the spin-orbit coupling lifts the degeneracy among the $2S+1$ spin states under zero magnetic field, leading to the zero-field splitting and magnetic anisotropy. 
If the $|m_s = \pm S\rangle$ states are the ground states after zero-field splitting, then the magnetic molecule gains a permanent magnetic moment at sufficiently low temperatures. This is a key feature of single-molecule magnets (SMMs), which may have data storage applications due to their bistability. 
The lowest $|m_s = \pm S\rangle$ states of SMMs form the basis for their applications as qubits. 
In general, regardless of the type of magnetic anisotropy of the molecule, the $n$ lowest spin states with $n>2$ form the basis for applications of the molecule as a qudit, as long as these states are energetically separated from higher excited states. 
Issues such as spin decoherence~\cite{Shiddiq2016-HoW10,Stewart2024-HoLF}, coupling of different qubits~\cite{Ghosh2021-Mn3dimer,Gakiya-Teruya2025-Hopzdo4}, and integration of molecular qubits into devices are under active investigation~\cite{Fursina2023-device,Chiesa2024-device}.
The above-mentioned properties make magnetic molecules promising candidates for applications in spintronics, magnetic storage, sensing, and quantum information processing.

%
Molecular spins in molecular crystals or on surfaces are open quantum systems due to their inevitable coupling to lattice vibrations.
The spin-phonon coupling leads to relaxation and decoherence of the spin states~\cite{Kragskow2023SpinPhonon}. It governs the rate at which the magnetization relaxes toward thermal equilibrium and may limit quantum coherence times in molecular spin qubits~\cite{Lunghi2022ExactPredictions,lunghi2026unified}.
In principle, the dynamics of the spin system is governed by the Liouville--von Neumann equation for the coupled spin-phonon system. 
In the weak-coupling limit, invoking the Born--Markov approximation and tracing out the phonon degrees of freedom yields the Redfield equation~\cite{Kragskow2023SpinPhonon} for the reduced density matrix, which describes the coupled time evolution of both spin populations and coherences.
An alternative at the same level of approximation is the generalized Lindblad equation of Saito \textit{et al.}~\cite{Saito2000QME}, which has been successfully applied to study the magnetization process of a nanoscale iron cluster~\cite{Nakano2001Fe6}. A reduced effective Hamiltonian for the large spin system of the iron cluster was constructed and adopted for computational feasibility. 
A further secular (rotating-wave) approximation yields an equation in the Gorini--Kossakowski--Sudarshan--Lindblad (GKSL) form~\cite{Breuer2002OpenQS}, commonly referred to as the secular Redfield equation.
Unlike the Redfield and generalized Lindblad equations, the GKSL equation is guaranteed to generate a completely positive, trace-preserving dynamical map.
Under a time-varying external magnetic field---as in pulsed-field magnetometry---the spin Hamiltonian becomes explicitly time-dependent and the system is driven out of thermal equilibrium in the presence of spin-phonon coupling, making the quantum master equation approach essential for capturing the resulting nonequilibrium magnetization dynamics~\cite{Rousochatzakis2005PulsedField}.

Several software packages have been developed to calculate the magnetic properties of molecular spin systems based on the spin Hamiltonian formalism. 
PHI, introduced by Chilton \textit{et al.}, is a widely used program that calculates equilibrium magnetic properties---including magnetization and magnetic susceptibility---and simulates electron paramagnetic resonance (EPR) spectra for mononuclear and polynuclear d- and f-block complexes~\cite{Chilton2013PHI}.
EasySpin is a MATLAB toolbox designed for simulating and fitting a broad range of continuous-wave and pulsed EPR spectra~\cite{Stoll2006EasySpin}.
Neither of them is designed for the time evolution of the density matrix under a varying magnetic field.
In this work, we present a computational utility named \textit{qdmag}~\cite{qdmag_code} for calculating the magnetization of magnetic molecules in the presence of spin-phonon coupling under a varying external magnetic field.
\textit{qdmag} adopts the generalized Lindblad equation of Saito \textit{et al.}~\cite{Saito2000QME}, but generalizes it in the ways required of a general-purpose utility.
It admits arbitrary exchange anisotropy together with extended Stevens operators to twelfth order in place of a fixed model Hamiltonian, it propagates the density matrix over millisecond timescales by piecewise-constant exponentiation of the Liouvillian, and it provides an alternative effective Hamiltonian suited to medium-sized systems dominated by isotropic exchange.

The related software addresses different observables. 
MolForge~\cite{Lunghi2022ExactPredictions} computes spin--phonon relaxation time from first principles through Redfield-type perturbation theory at fixed field. It is complementary to \textit{qdmag} rather than overlapping with it. MolForge predicts the rates that a given spin--phonon coupling produces. In contrast, \textit{qdmag} takes the coupling as a phenomenological spectral density and returns the nonequilibrium magnetization $M(t)$ under an arbitrary field profile.
STOSS~\cite{GutierrezFinol2023} simulates the magnetization dynamics of molecular nanomagnets through a classical Markov-chain Monte Carlo evolution of Boltzmann-weighted transition probabilities. It does not propagate a density matrix, and therefore does not capture coherent features such as the Rabi-type oscillation reported in Section~\ref{sec:case1} and \ref{sec:HoAna}.
General Liouville-space engines such as Spinach~\cite{Hogben2011} and QuTiP~\cite{Johansson2013QuTiP} supply comparable propagation machinery, including the piecewise-constant exponential stepping that \textit{qdmag} also adopts. \rev{Spinach also provides a spin-Hamiltonian front end and a general Liouville-space framework in which new relaxation theories can be easily encoded. The generalized-Lindblad spin-phonon dissipator of Saito \textit{et al.}~\cite{Saito2000QME} and a staircase pulsed-field profile were implemented in an example set that reproduces the corresponding results in the present work~\cite{Spinach2026PulsedField}. \textit{qdmag} is a lightweight, dedicated Python alternative that differs from this implementation in two respects. First, by default it propagates each stair with the exact exponential of the full Liouvillian (Eq.~\ref{eq:prop}) rather than approximately via symmetric (Strang) operator splitting~\cite{Strang1968} and a second-order expansion of the dissipator. Second, it builds the effective basis of Section~\ref{sec:basis} internally, in a uniquely defined basis that also fixes the operator $X$.}

Other treatments of magnetization in a time-varying field occupy adjacent but distinct niches. 
F\"oldi \textit{et al.}~\cite{Foldi2007} solved the time-dependent Schr\"odinger equation with phase relaxation treated through a master equation. This resolved the hysteresis steps as sequences of multilevel avoided crossings beyond the Landau--Zener--St\"uckelberg picture, but without treating general polynuclear Hamiltonians or thermal spin--phonon dissipation on the same footing.
Rousochatzakis and Luban~\cite{Rousochatzakis2005PulsedField} formulated pulsed-field master equations in the adiabatic basis, combining Landau--Zener--St\"uckelberg transitions with thermal rates. That formulation reduces to population rate equations and does not retain the coherences that \textit{qdmag} propagates between level crossings.

\section{Spin Hamiltonian}

Transition metal and rare-earth complexes can host localized spins due to partially filled $d$ or $f$ orbitals. 
Their low-energy states can be described effectively by the spin Hamiltonian~\cite{Rudowicz2001SpinH} given in Eq.~\ref{eq:spinH}.
\begin{equation}
    \label{eq:spinH}
    \hat{H}(t) = \hat{H}_\text{ex} + \hat{H}_\text{ZFS} + \hat{H}_\text{Zee}(t)
\end{equation}

$\hat{H}_\text{ex}=-2\sum_{ij}\hat{\boldsymbol{S}}_i^\mathrm{T} \boldsymbol{J}_{ij} \hat{\boldsymbol{S}}_j$ is the magnetic exchange interaction, where $i$ and $j$ are the indices of the magnetic ions, $\hat{\boldsymbol{S}}$ (subscript omitted for brevity) is a column vector of the spin operators $\hat{S}_x$, $\hat{S}_y$, and $\hat{S}_z$, $\boldsymbol{J}$ is a $3 \times 3$ matrix of exchange coupling constants, and the superscript T denotes matrix or vector transpose. The exchange coupling matrix $\boldsymbol{J}$ can be decomposed into an isotropic part
\begin{equation}
\label{eq:Jiso}
\boldsymbol{J}^\text{iso} = 
\begin{pmatrix}
J^\text{iso} & 0 & 0 \\
0 & J^\text{iso} & 0 \\
0 & 0 & J^\text{iso}
\end{pmatrix}\text{,}
\end{equation}
a traceless anisotropic part 
\begin{equation}
\label{eq:Jani}
\boldsymbol{J}^\text{ani} = 
\begin{pmatrix}
J^\text{ani}_{xx} & 0 & 0 \\
0 & J^\text{ani}_{yy} & 0 \\
0 & 0 & J^\text{ani}_{zz}
\end{pmatrix}\text{,}
\end{equation}
an antisymmetric part 
\begin{equation}
\label{eq:Jasym}
\boldsymbol{J}^\text{asym} = 
\begin{pmatrix}
0 & J^\text{asym}_{xy} & J^\text{asym}_{xz} \\
-J^\text{asym}_{xy} & 0 & J^\text{asym}_{yz} \\
-J^\text{asym}_{xz} & -J^\text{asym}_{yz} & 0
\end{pmatrix}\text{,}
\end{equation}
and a symmetric part
\begin{equation}
\label{eq:Jsym}
\boldsymbol{J}^\text{sym} = 
\begin{pmatrix}
0 & J^\text{sym}_{xy} & J^\text{sym}_{xz} \\
J^\text{sym}_{xy} & 0 & J^\text{sym}_{yz} \\
J^\text{sym}_{xz} & J^\text{sym}_{yz} & 0
\end{pmatrix}\text{.}
\end{equation}
In the above equations, the isotropic exchange coupling constant $J^\text{iso}$ equals $\text{Tr}(\boldsymbol{J})/3$. 
The diagonal matrix elements of the anisotropic part are determined by $\boldsymbol{J}^\text{ani}_{\alpha\alpha}=J_{\alpha\alpha} - J^\text{iso}$, where $\alpha=x,y,\, \text{and}\, z$. $\boldsymbol{J}^\text{ani}$ is traceless according to the definition, and thus it only has two degrees of freedom. 
All diagonal matrix elements of both $\boldsymbol{J}^\text{asym}$ and $\boldsymbol{J}^\text{sym}$ are zero, and their off-diagonal matrix elements equal the corresponding ones of $(\boldsymbol{J} - \boldsymbol{J}^\text{T})/2$ and $(\boldsymbol{J} + \boldsymbol{J}^\text{T})/2$, respectively. 
The sum of $\boldsymbol{J}^\text{iso}$ and $\boldsymbol{J}^\text{ani}$ recovers the diagonal part of the complete exchange coupling matrix. Meanwhile, the sum of $\boldsymbol{J}^\text{asym}$ and $\boldsymbol{J}^\text{sym}$ recovers the off-diagonal part.

In general, the zero-field splitting (ZFS) term can be written as
\begin{equation}
    \label{eq:zfs}
    \hat{H}_\text{ZFS}=\sum_{i,kq} B_{i,k}^q \hat{O}_{i,k}^q(S_i)
\end{equation}
where $\hat{O}_{i,k}^q(S_i)$ are the extended Stevens operators ~\cite{Hutchings1964PointCharge,Rudowicz2004ESO,Stoll2006EasySpin} for the effective spin $S_i$ of the $i$th magnetic ion and $B_{i,k}^q$ are ZFS parameters. Here, $k$ specifies the order of the extended Stevens operator and takes the values $2,4,\cdots,2S$ for the effective spin $S$.~\cite{Chibotaru2012SINGLEANI} $q$ specifies a component of the extended Stevens operator and takes the values $-k, -k+1, \cdots, k$. For lanthanide ions, the total angular momentum provides a suitable choice for the effective spin due to strong spin-orbit coupling. 

Lastly, the Zeeman term, given by
\begin{equation}
    \label{eq:zee}
    \hat{H}_\text{Zee}(t)= \sum_i \mu_B \boldsymbol{B}^\mathrm{T}(t) \mathbf{g}_i \boldsymbol{S}_i,
\end{equation}
describes the interaction between the local spins and the time-dependent external magnetic field. Here, $\boldsymbol{g}$ is the Land\'{e} g-tensor and $\mu_B$ is the Bohr magneton. Without losing generality, we choose the axis that is parallel with the magnetic field to be the $z$-axis in our numerical implementation. The direction of the external magnetic field is assumed to remain unchanged throughout the sweep.

\section{Quantum master equation} 

Previously, a quantum master equation in the generalized Lindblad form was derived for an open quantum system coupled to a phonon bath under the assumption of weak coupling~\cite{Saito2000QME}. It has been successfully applied to understand the magnetization processes of a \ce{Fe6}~\cite{Nakano2001Fe6} magnetic molecule.
The quantum master equation reads
\begin{equation}
    \label{eq:qme}
    \frac{\textrm{d}\rho(t)}{dt} = \frac{1}{i\hbar}[\hat{H}(t), \rho(t)] - \Gamma\rho(t).
\end{equation}
In Eq.~\ref{eq:qme}, $\hat{H}(t)$ is the spin Hamiltonian in Eq.~\ref{eq:spinH} for the magnetic molecule, and the $\Gamma\rho(t)$ term accounts for the spin-phonon coupling. In particular,
\begin{equation}
\label{eq:Grho}
\Gamma\rho(t) = \{[X, R\rho(t)] + [X, R\rho(t)]^{\dagger}\}\lambda^2\pi/\hbar
\end{equation}
where $\lambda$ is the spin-phonon coupling constant. $X$ is an operator in the spin space with the matrix elements~\cite{Nakano2001Fe6}
\begin{equation}
\label{eq:X}
\langle m_s | X | m_s^\prime \rangle = 
\begin{cases}
  1 & \text{if } |m_s - m_s^\prime|=1 \\
  0 & \text{otherwise}
\end{cases}
\end{equation}
in the basis of the eigenstates of the total $\hat{S}_z$ operator. \rev{For a single spin, the basis is the standard $|S, m_s\rangle$ basis, and the operator $X$ is uniquely defined. For a multi-spin system, the eigenvalues of $\hat{S}_z$ can be degenerate, and $X$ is no longer unique. To define $X$ uniquely in the latter case, \textit{qdmag} takes the basis states to be the common eigenstates of the isotropic exchange term, the total $\hat{S}_z$, and the squared total spin operator $\hat{S}^2$. Multiplets of equal energy and equal $S$, if any, are told apart by the intermediate-spin operators $(\hat{S}_1 + \hat{S}_2)^2$, $(\hat{S}_1 + \hat{S}_2 + \hat{S}_3)^2$, $\ldots$. The phases are fixed by the Condon--Shortley convention, with the largest component of each $m_s = S$ state real and positive.} In the basis of the instantaneous eigenstates of the spin Hamiltonian $\hat{H}(t)$, the matrix elements of the $R$ operator are 
\begin{equation}
    \label{eq:R}
    \langle k|R|n \rangle = \frac{1}{\hbar} X_{kn} \Phi \left( \frac{E_k - E_n}{\hbar} \right)
\end{equation}
where 
\begin{equation}
    \label{eq:Phi}
    \Phi(\omega) = \frac{I(\omega) - I(-\omega)}{\mathrm{e}^{\beta \hbar \omega} - 1}.
\end{equation}
In Eq.~\ref{eq:Phi}, $\beta=1/k_B T$ with $k_B$ the Boltzmann constant and $T$ the temperature of the phonon bath. It is noteworthy that a consistent representation of the $X$ and $R$ operators needs to be used in numerical calculations. In Eq.~\ref{eq:Phi}, the spectral density $I(\omega)$ as a function of the phonon frequency depends on the phonon density of states as well as the mode-dependent spin-phonon coupling.~\cite{Saito2000QME} We implement the following form of the spectral density
\begin{equation}
    \label{eq:I}
    I(\omega) = I_0 \omega^\alpha \theta(\omega)
\end{equation}
where $\theta(\omega)$ is the step function. The phonon bath is called sub-Ohmic, Ohmic, and super-Ohmic for $\alpha<1$, $\alpha=1$, and $\alpha>1$, respectively~\cite{Grabert1988QBrown}. In our code, $\alpha$ defaults to 2 but can be adjusted by the user. \rev{Dimensional consistency of Eqs.~\ref{eq:Grho}--\ref{eq:Phi} requires $I(\omega)$ to carry the unit of time. With $\omega$ in $\textrm{rad}\cdot\textrm{ps}^{-1}$ and $I(\omega)$ in ps, the prefactor $I_0$ carries the unit $\textrm{ps}^{\alpha+1}\cdot\textrm{rad}^{-\alpha}$, i.e., $\textrm{ps}^{3}\cdot\textrm{rad}^{-2}$ for the default $\alpha=2$.} 
Besides $\alpha$, the only other adjustable parameter for the $\Gamma\rho(t)$ term in the quantum master equation is the prefactor $I_0 \lambda^2$. Different combinations of $I_0$ and $\lambda$ yield the same results when $I_0 \lambda^2$ remains constant. Therefore, we fixed $\lambda$ at $10\; \mathrm{cm}^{-1}$ and varied $I_0$ in the case studies presented in Section~\ref{sec:cases}. 
In Eq.~\ref{eq:qme}, $\rho(t)$ is the reduced density matrix for the spin system, and the dynamics governed by Eq.~\ref{eq:qme} preserves its conditions of normalization, Hermiticity, nonnegativity, and the Cauchy-Schwarz inequality, as given in \ref{sec:conditions}.

\subsection{The Liouville form}

The Liouville form of quantum master equations is compact and convenient for numerical calculations~\cite{Campaioli2024QME}. Due to the Hermitian conjugate term in Eq.~\ref{eq:Grho}, it is necessary to separate the real and imaginary parts of the density matrix $\rho$ in order to rewrite Eq.~\ref{eq:qme} in the Liouville form. 
To this end, the matrix of $\rho$ in any basis is first flattened into a vector of complex numbers using the row-major ordering. The imaginary part of this vectorized density matrix is appended to the vector of the real part, doubling the length of the vector. In what follows, we abuse the notation $\rho$ by using it to denote the resulting vector of real numbers with dimension $2n^2$, where $n$ is the dimension of the spin space. 
The Liouville form of Eq.~\ref{eq:qme} is 
\begin{equation}
    \label{eq:liouville}
    \frac{\text{d}}{\text{d}t} \rho = \mathcal{L} \rho 
\end{equation}
where the Liouvillian $\mathcal{L}$ is
\begin{equation}
    \label{eq:liouvillian}
    \mathcal{L} = 
    \begin{pmatrix}
        \mathcal{L}_{11} & \mathcal{L}_{12} \\
        \mathcal{L}_{21} & \mathcal{L}_{22} 
    \end{pmatrix}
\end{equation}
with 
\begin{equation}
    \label{eq:L11}
    \mathcal{L}_{11} = \frac{1}{\hbar}\text{Im}(A) - \frac{\lambda^2 \pi}{\hbar} \left(\text{Re}(B) + \text{Re}(B^\prime)\right), 
\end{equation}
\begin{equation}
    \label{eq:L12}
    \mathcal{L}_{12} = \frac{1}{\hbar}\text{Re}(A) + \frac{\lambda^2 \pi}{\hbar} \left(\text{Im}(B) + \text{Im}(B^\prime)\right), 
\end{equation}
\begin{equation}
    \label{eq:L21}
    \mathcal{L}_{21} =-\frac{1}{\hbar}\text{Re}(A) - \frac{\lambda^2 \pi}{\hbar} \left(\text{Im}(B) - \text{Im}(B^\prime)\right), 
\end{equation}
and
\begin{equation}
    \label{eq:L22}
    \mathcal{L}_{22} = \frac{1}{\hbar}\text{Im}(A) - \frac{\lambda^2 \pi}{\hbar} \left(\text{Re}(B) - \text{Re}(B^\prime)\right). 
\end{equation}
In Eqs.~\ref{eq:L11}--\ref{eq:L22}, the matrix elements of $A$ and $B$ are
\begin{equation}
    \label{eq:A}
    A_{in+j,kn+l}=H_{ik}\delta_{lj} - \delta_{ik}H_{lj},
\end{equation}
and
\begin{equation}
    \label{eq:B}
    B_{in+j,kn+l}=(XR)_{ik} \delta_{lj} - R_{ik}X_{lj}.
\end{equation}
$B^\prime$ is related to $B$ by $B_{in+j,kn+l}^\prime=B_{jn+i,kn+l}$. The indices $i$, $j$, $k$, and $l$ run from $1$ to $n$.

\subsection{The effective basis} 
\label{sec:basis}

The dimension of spin space for a multinuclear complex is $N = \Pi_i (2S_i +1)$ which could be so large that (repeated) direct diagonalization of the spin Hamiltonian becomes computationally infeasible or unnecessary. Intuitively, thermally unoccupied eigenstates of the spin Hamiltonian during the whole dynamics are not necessary for magnetization processes. In these cases an effective Hamiltonian of a smaller dimension than $N$ is desirable. 
Irrespective of the type of magnetic interactions in the multinuclear complex, the saturated magnetic moment of the system under high magnetic field is determined by the maximal total spin, which is $S_\text{max}=\sum_i S_i$. To capture all possible quantized magnetization plateaus across all magnetic fields at low temperatures, a minimal effective Hamiltonian should span all the spin states $m_s = -S_\text{max}, -S_\text{max}+1, \ldots, S_\text{max}$ and thus has a dimension of $2S_\text{max}+1$.
A method for devising such an effective Hamiltonian has been proposed by Nakano and Miyashita~\cite{Nakano2001Fe6} for large systems for which direct diagonalization is not feasible. In the following, we present an alternative method for medium-sized systems based on direct diagonalization. The method can reproduce the target spin states faithfully and is useful for systems dominated by the isotropic exchange interaction.

First, the basis states for the effective Hamiltonian are chosen as follows. 
A Zeeman term $\hat{H}_\text{Zee}$ with a small magnetic field ($10^{-4}$--$10^{-3}$ T) along the $z$ direction and an isotropic Land\'{e} g-tensor for each spin is added to the isotropic exchange term $\hat{H}_\text{ex}$. The operator sum $\hat{H}_\text{ex}+\hat{H}_\text{Zee}$ is then diagonalized, yielding the common eigenstates of the isotropic exchange term and the total $S_z$ operator. This is true since the perturbative Zeeman term commutes with the isotropic exchange term as well as the total $S_z$ operator. \rev{The collection of all these eigenstates will be referred to as the pool of basis states, which are made unique by the aforementioned intermediate-spin operators and phase convention.} 
For each $z$-projection of the total spin $m_s$, the lowest-energy state is picked among all eigenstates of $\hat{H}_\text{ex}+\hat{H}_\text{Zee}$ with the same $m_s$. This leads to $2S_\text{max}+1$ states which can serve as the basis set for the minimal effective Hamiltonian. More eigenstates of $\hat{H}_\text{ex}+\hat{H}_\text{Zee}$ can be picked and included in the basis set to improve the effective Hamiltonian. 
It is noteworthy that the eigenstates of the isotropic exchange term alone as obtained by direct diagonalization are often not the eigenstates of the total $S_z$ operator. This is why the perturbative Zeeman term is used in the above numerical procedure.

Second, the effective Hamiltonian is constructed with the chosen basis states $|1\rangle, |2\rangle, \ldots,\text{and } |n\rangle$. The matrix elements of the effective Hamiltonian are $\hat{H}_{ij}^\text{eff} =\langle i | \hat{H} | j \rangle$, where $\hat{H}$ is the full spin Hamiltonian in Eq.~\ref{eq:spinH}. Apparently, the dimension of the effective Hamiltonian is $n \times n$, which can be much smaller than the dimension of the full spin Hamiltonian. The effective $X$, $R$, and $\hat{\boldsymbol{S}}$ operators are constructed in the same way. Eventually, the entire quantum master equation in Eq.~\ref{eq:qme} is written in terms of the chosen basis states and solved in this representation.

It is noteworthy that the $R$ operator depends on time implicitly. The time dependence of $R$ arises from the eigenvalues of the spin Hamiltonian, which vary with time since the Zeeman term is time-dependent. In numerical calculations, the matrix form of the $R$ operator is first obtained on the basis of the eigenstates of the effective Hamiltonian according to the definition in Eq.~\ref{eq:R} at each time step. A subsequent change of representation is performed to cast the $R$ operator onto the basis states $|i\rangle$, $i=1,2,\ldots,n$. The effective Zeeman term can be constructed efficiently by substituting the spin operators in Eq.~\ref{eq:zee} with the effective spin operators, which are time independent. The effective spin operators and all other time-independent operators only need to be constructed once at the beginning of the time evolution. 

\subsection{The staircase approximation} 

\begin{figure}[htb!]
\centering
\includegraphics[width=0.6\columnwidth]{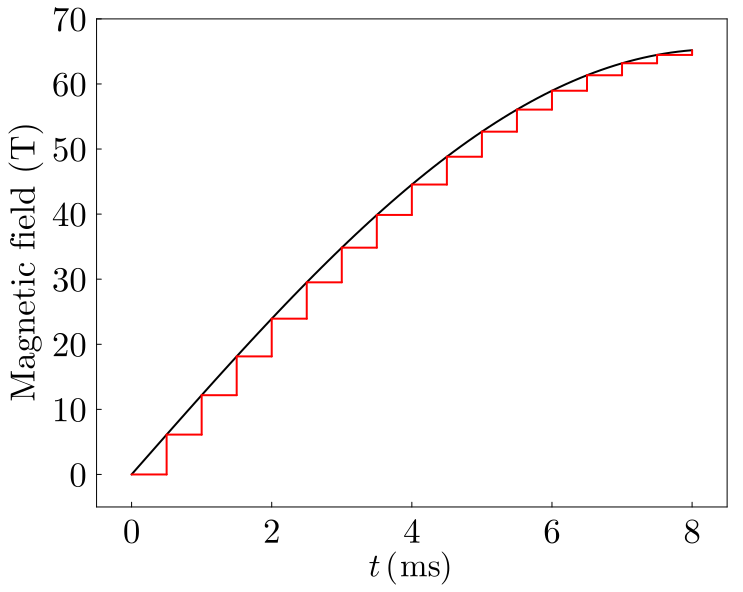}
\caption{\label{fig:staircase}Illustration of the staircase function used to approximate the continuous variation of magnetic field with time.}
\end{figure}

At a constant magnetic field, the whole spin Hamiltonian in Eq.~\ref{eq:spinH} is time-independent. In this case, the quantum master equation has the following solution~\cite{Blanes2009Magnus}
\begin{equation}
    \label{eq:prop}
    \rho(t+\Delta t)=\textrm{exp}(\mathcal{L} \Delta t) \rho(t).
\end{equation}
It is noteworthy that the solution in Eq.~\ref{eq:prop} is exact for both short and long time steps $\Delta t$. This inspires us to implement the staircase approximation where the magnetic field profile versus time is replaced with a staircase function as illustrated in Fig.~\ref{fig:staircase}.
The quality of the staircase approximation is controlled by the width of each stair which should be reduced to converge the calculated magnetization. 
As demonstrated in \ref{sec:RK4}, the staircase approximation maintains numerical stability with much, and sometimes orders-of-magnitude, larger time steps compared to the Runge–Kutta method, allowing for more efficient simulations with longer time evolution periods.
The reason is as follows. Because Eq.~\ref{eq:prop} gives the exact solution for the system subject to the staircase-approximated field profile, the staircase approximation reproduces the magnetization evolution accurately as long as the system cannot distinguish the staircase profile from the original smooth one. This holds when the time step is much shorter than both the timescale over which the field varies appreciably and the timescales of the system's response: the population-transfer processes such as spin-phonon relaxation, Landau-Zener transitions, and Rabi oscillations. The Runge-Kutta method, on the other hand, computes the magnetization dynamics by accurately resolving the evolution of the system's quantum state, including the fast-oscillating coherences between states. Consequently, both its stability and its accuracy require the time step to be much smaller than the inverse of the largest-magnitude eigenvalue of the Liouvillian. This eigenvalue is typically on the order of the largest energy gap of the Hamiltonian and corresponds to the fastest timescale of the problem.
\textit{qdmag} adopts three different methods for evaluating the product $\textrm{exp}(\mathcal{L} \Delta t) \rho$ in Eq.~\ref{eq:prop}. The Pad\'{e} method is the most efficient at long time steps (see \ref{sec:propagators}), and thus it is used throughout this work.

\section{Case studies} 
\label{sec:cases}

This section presents three case studies to demonstrate the use of the \textit{qdmag} code. The first, second, and third cases deal with a single-spin system, a two-spin system, and a three-spin system, respectively. The full Hilbert space was used to solve the quantum master equation for the first two cases due to the small system size. A reduced Hilbert space was applied in the last case. \rev{The input files of all case studies are distributed with the code, and the input file format is described in \ref{sec:input}.} 

\subsection{A single spin of \texorpdfstring{$J=8$}{J=8}} 
\label{sec:case1}

A Ho-based metal organic framework \ce{[Ho(pyrazine-1,4-dioxide)4](ClO4)3} (\ce{Ho(pzdo)4})~\cite{Gakiya-Teruya2025-Hopzdo4} was recently reported to exhibit a clock transition gap which enhances spin coherence~\cite{Shiddiq2016-HoW10}. \ce{Ho(pzdo)4} has localized spins of a total angular momentum $J=8$ around the \ce{Ho^{3+}} ions due to the partially occupied $4f$ shell. The low-energy spin excitations for each \ce{Ho^{3+}} ion can be described by the following effective spin Hamiltonian.
\begin{equation}
    \label{eq:HHo}
    \hat{H} = \sum_{k=2,4,\ldots,12} \sum_{q=-k}^k B_k^q \hat{O}_k^q(J) + g_\textrm{eff} \mu_B B(t) \hat{J}_z
\end{equation}
In Eq.~\ref{eq:HHo}, the $g_\textrm{eff}$ factor is set to $1.24$ and the ZFS parameters are taken from literature~\cite{Gakiya-Teruya2025-Hopzdo4}. Both the effective $g$-factor and the ZFS parameters were calculated using the complete active space self-consistent field method. As aforementioned, the external magnetic field is assumed to be along the $z$-axis in the Zeeman term. An arbitrary molecular orientation is dealt with by transforming the extended Stevens operators $\hat{O}_k^q$ from the tensor frame for the ZFS parameters to the lab frame. All eigenstates of the $\hat{J}_z$ operator are used as basis states to solve the quantum master equation. 

\begin{figure}[htb!]
\centering
\includegraphics[width=0.95\columnwidth]{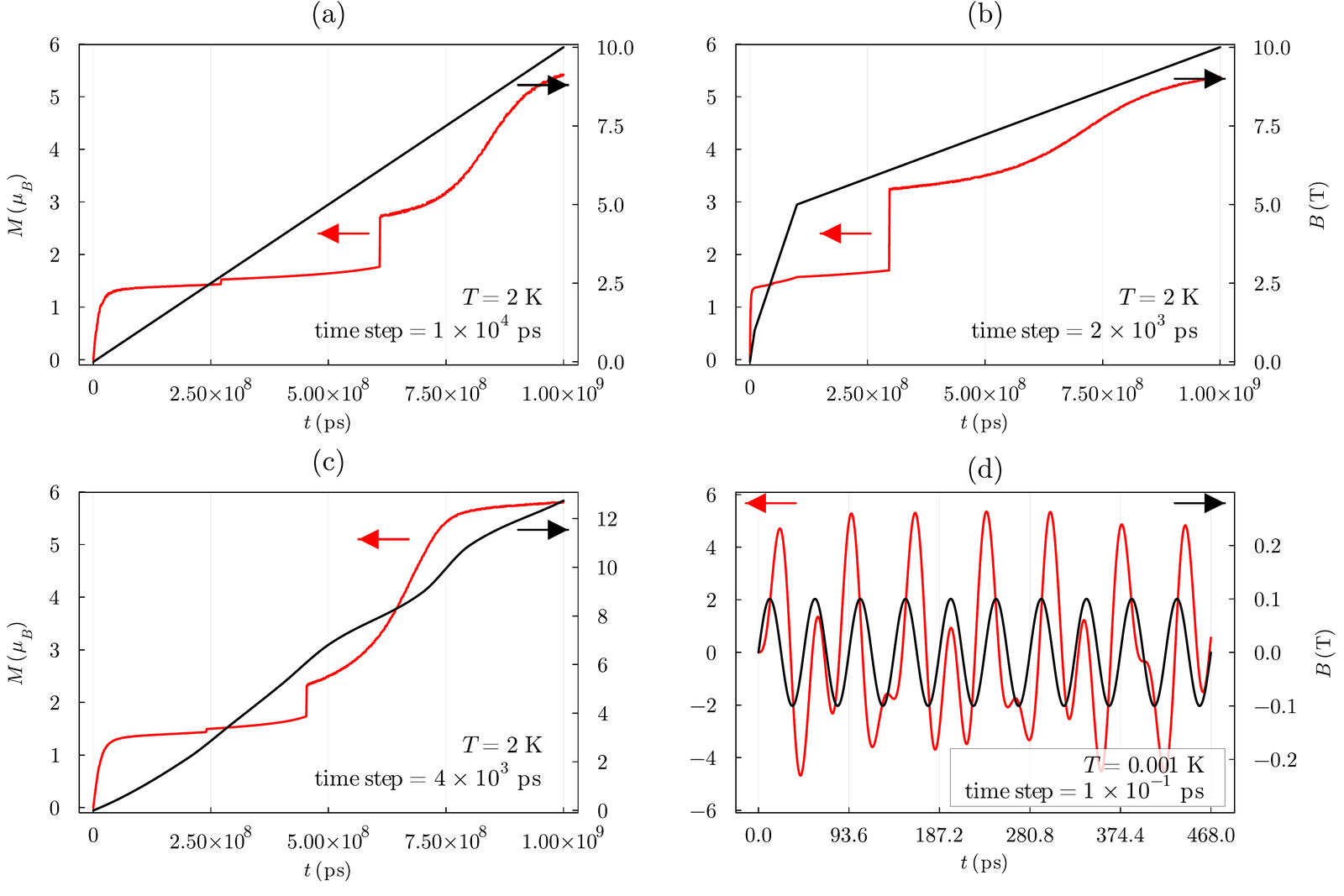}
\caption{\label{fig:1_MB_pulse}Four different magnetic field profiles and the corresponding magnetization versus time for \ce{Ho(pzdo)4}. Parameters used in calculations: $\mathrm{sweep\; rate}=10~\mathrm{T/ms}$ for the linear function in the panel (a); Turning points $(1~\mu\mathrm{s},0.1~\mathrm{T})$, $(10~\mu\mathrm{s},1~\mathrm{T})$, $(100~\mu\mathrm{s},5~\mathrm{T})$, and $(1000~\mu\mathrm{s},10~\mathrm{T})$ for the piecewise linear function in the panel (b);  The cubic spline in the panel (c) is fitted to the experimental data\protect\footnotemark; An amplitude of $0.1~\mathrm{T}$ and a period of $46.8~\mathrm{ps}$ for the sinusoidal function in the panel (d). The temperature and time step are marked in the bottom right corner of each panel. \rev{The time steps of panels (b) and (c) are chosen such that the field step per stair at the avoided crossings does not exceed that of panel (a).} $I_0 = 1\times 10^{-14}~{\textrm{ps}^{3}\cdot \textrm{rad}^{-2}}$ is used in all calculations.}
\end{figure}

The \textit{qdmag} code accepts the following four types of magnetic field as a function of time: 1) A linear function of time $B(t) = a t$, where $a$ is the sweep rate; 2) A piecewise linear function of time; 3) A cubic spline fit to the experimental magnetic pulse data; and 4) A sinusoidal function $B(t) = B_0 \mathrm{sin}(\omega t)$, where $B_0$ is the amplitude and $\omega$ is the angular frequency. 
All these four types of $B(t)$ are applied to \ce{Ho(pzdo)4} in our numerical calculations. 
The resulting magnetization curves together with the corresponding magnetic field profiles are shown in Figs.~\ref{fig:1_MB_pulse}a-\ref{fig:1_MB_pulse}d. The temperature is $2~\textrm{K}$ for the results in Figs.~\ref{fig:1_MB_pulse}a--\ref{fig:1_MB_pulse}c and $0.001~\textrm{K}$ for the results in Fig.~\ref{fig:1_MB_pulse}d. 
\footnotetext{As produced by the 65 Tesla short pulse magnet of the National High Magnetic Field Laboratory. It is available on the website nationalmaglab.org.}
The first three $B(t)$ functions in Figs.~\ref{fig:1_MB_pulse}a--\ref{fig:1_MB_pulse}c share a similar average sweep rate but differ in the local sweep rates. The resulting magnetization curves share similar features such as the highest magnetic moment between 5 and 6 $\mu_B$ at $1~\textrm{ms}$, a magnetization plateau between 1 and 2 $\mu_B$, and a sudden increase that follows the plateau. Varying the local sweep rate modulates the slope of the magnetization curve and influences the time at which the sudden increase of magnetization occurs. The linear and spline-fitted $B(t)$ profiles additionally produce a secondary sudden increase in magnetization as well as a secondary magnetization plateau preceding the common magnetization plateau. As shown in \ref{sec:HoAna}, the sudden jumps in the magnetization curve are due to \rev{avoided crossings of the Zeeman energy levels}. 
For the last sinusoidal function $B(t)$ in Fig.~\ref{fig:1_MB_pulse}d, the angular frequency $\omega$ is set such that $\hbar \omega$ matches the calculated clock transition gap of $0.71~\textrm{cm}^{-1}$,\footnote{The measured clock transition gap is $1.82~\textrm{cm}^{-1}$ or $54.6~\textrm{GHz}.$} and the amplitude $B_0$ is set to $0.1~\textrm{T}$. In this case, the magnetization oscillates non-periodically between $-4.68\,\mu_B$ and $+5.36\,\mu_B$. Such a magnetization oscillation is a synergy of Rabi oscillation and state composition modulation due to the applied oscillating magnetic field, see \ref{sec:HoAna} for detailed analysis.
Since the temperature is as low as $0.001~\textrm{K}$, the effects of spin-phonon coupling are negligible and the solution is essentially the same as the coherent solution to the von Neumann equation, i.e. Eq.~\ref{eq:qme} without the $\Gamma \rho$ term. 
The time steps for all four cases are set sufficiently small to guarantee convergence of the magnetization. The convergence test results are given in \ref{sec:convergence}.

\begin{figure}[htb!]
\centering
\includegraphics[width=0.6\columnwidth]{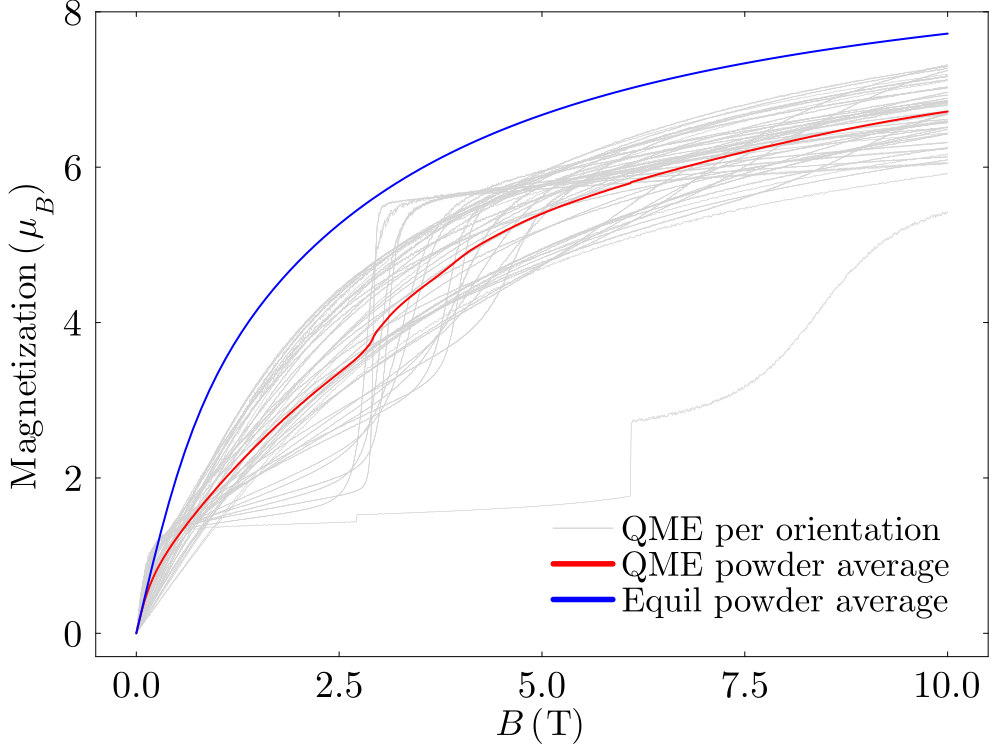}
\caption{\label{fig:1_MB_powder}Magnetization versus magnetic field before and after powder average. The equilibrium magnetization was given for comparison. Parameters used in calculations: $T=2\,\textrm{K}$, $I_0 = 1\times 10^{-14}\,{\textrm{ps}^{3}\cdot \textrm{rad}^{-2}}$, sweep rate $=10\,\mathrm{T/ms}$, and time step $1\times 10^4\,\textrm{ps}$.}
\end{figure}

The magnetization along the external magnetic field depends on the orientation of the system due to  magnetic anisotropy. In the above calculations, the ZFS tensor frame is aligned with the lab frame. For powder samples, the net magnetization is an average over the magnetization of all powder grains which have random orientations. In numerical calculations, the powder average can be calculated as follows. 
\begin{equation}
\label{eq:powder}
\begin{aligned}
\overline{M} &= \frac{1}{8\pi^2}\int_{0}^{2\pi} \mathrm{d}\alpha \int_{0}^{\pi} \sin(\beta)\mathrm{d}\beta \int_{0}^{2\pi} \mathrm{d}\gamma M(\alpha, \beta, \gamma) \\
         &= \frac{1}{4\pi}\sum_{i,j} w^l_{ij}M(0, \beta_{ij}, \gamma_{ij})
\end{aligned}
\end{equation}
In Eq.~\ref{eq:powder}, $M(\alpha, \beta, \gamma)$ is the magnetization along the lab $z$-axis when the system is rotated actively using the Euler angles $\alpha$, $\beta$, and $\gamma$. Note that the subscripts for the Euler angles are omitted for brevity. The rotation is intrinsic and follows the $zyz$ convention. The angle $\alpha$ is applied first, followed by $\beta$ and $\gamma$. 
In the second line of Eq.~\ref{eq:powder}, $\alpha$ is integrated out, yielding a factor of $2\pi$, since different values of $\alpha$ correspond to the same magnetization.
$w^l_{ij}$ are the weights for the sampled points using the Lebedev quadrature. The quadrature points and the associated weights are obtained by the SciPy package.~\cite{2020SciPy-NMeth} A Lebedev order of 21 is adopted in our calculations, leading to 170 system orientations for each of which the quantum master equation is solved independently. We note that the numerical error in the powder-averaged magnetization is within $0.01\, \mu_B$ according to test calculations in thermal equilibrium (see \ref{sec:powderconv}). The calculated magnetization curves for \ce{Ho(pzdo)4} under the linear magnetic field profile are shown as the gray lines in Fig.~\ref{fig:1_MB_powder}. Depending on the system orientation and the magnetic field, the calculated magnetization can differ by over $3\, \mu_B$. After powder averaging, the kinks in the magnetization curve become less pronounced but remain noticeable. Compared with the equilibrium magnetization, the magnetization obtained by solving the quantum master equation is lower throughout the entire magnetic field range from 0 to 10 T. This is because the thermal dissipation into the phonon bath is not quick enough to thermalize the spin system. With a larger spin-phonon coupling, the nonequilibrium magnetization curve would approach the equilibrium one. 

\subsection{Two coupled spins of \texorpdfstring{$S=1/2$}{S=1/2}} 

Next, we examine a toy model where two spins of $S=1/2$ are coupled by magnetic exchange coupling. Such a model could be suitable for describing transition metal dimers with local spins of $S=1/2$ at the metal centers. The spin Hamiltonian for this toy model is as follows. 
\begin{equation}
\hat{H} = -2 \hat{\boldsymbol{S}}_1^T \boldsymbol{J}^\textrm{dimer} \hat{\boldsymbol{S}}_2 + g_\textrm{s} \mu_B B(t) (\hat{S}_{1,z} + \hat{S}_{2,z})
\end{equation}
The dimension of the Hilbert space for this model is 4. All four \rev{pool states} are used to solve the quantum master equation. Four kinds of exchange coupling as given in Eqs.~\ref{eq:Jdimer1}--\ref{eq:Jdimer4} are considered. 
\begin{equation}
\label{eq:Jdimer1}
\boldsymbol{J}^\textrm{dimer}_1 = 
\begin{pmatrix}
0.20 &  0.00 &  0.00 \\
0.00 &  0.20 &  0.00 \\
0.00 &  0.00 &  0.20
\end{pmatrix}\; \textrm{cm}^{-1}
\end{equation}
\begin{equation}
\label{eq:Jdimer2}
\boldsymbol{J}^\textrm{dimer}_2 = 
\begin{pmatrix}
0.20 &  0.00 &  0.00 \\
0.00 &  0.00 &  0.00 \\
0.00 &  0.00 &  0.00
\end{pmatrix}\; \textrm{cm}^{-1}
\end{equation}
\begin{equation}
\label{eq:Jdimer3}
\boldsymbol{J}^\textrm{dimer}_3 = 
\begin{pmatrix}
0.00 &  0.00 &  0.00 \\
0.00 &  0.00 &  0.00 \\
0.00 &  0.00 &  0.20
\end{pmatrix}\; \textrm{cm}^{-1}
\end{equation}
\begin{equation}
\label{eq:Jdimer4}
\boldsymbol{J}^\textrm{dimer}_4 = 
\begin{pmatrix}
 0.00 &  0.20 &  0.20 \\
-0.20 &  0.00 &  0.20 \\
-0.20 & -0.20 &  0.00
\end{pmatrix}\; \textrm{cm}^{-1}
\end{equation}
\begin{figure}[htb!]
\centering
\includegraphics[width=0.6\columnwidth]{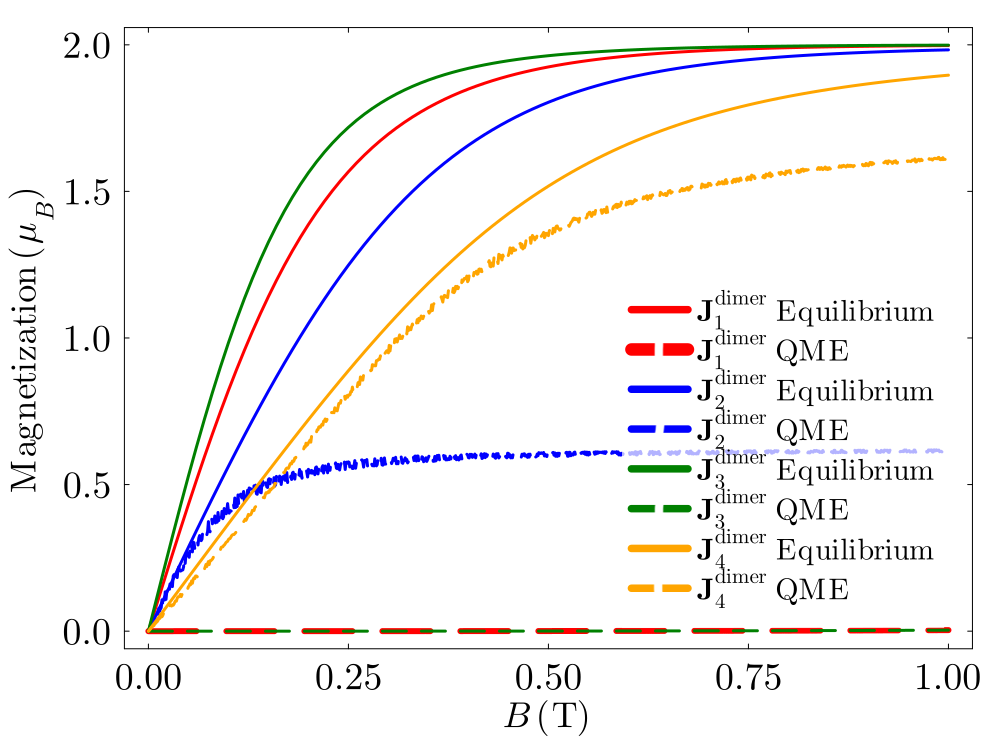}
\caption{\label{fig:2_MB_J}Magnetization versus magnetic field using different exchange coupling tensors for a dimer. Parameters used in calculations: $T=0.2~\textrm{K}$, $I_0 = 1\times 10^{-10}~{\textrm{ps}^{3}\cdot \textrm{rad}^{-2}}$, sweep rate $=10~\textrm{T}/\textrm{ms}$, and time step $=1\times 10^4~\textrm{ps}$.}
\end{figure} 
Among the above four exchange coupling matrices, $\boldsymbol{J}^\textrm{dimer}_1$ is isotropic, $\boldsymbol{J}^\textrm{dimer}_2$ and $\boldsymbol{J}^\textrm{dimer}_3$ are anisotropic, and $\boldsymbol{J}^\textrm{dimer}_4$ is antisymmetric. For the Zeeman term, the magnetic field is applied along the $z$-axis, and the spin $g$-factor $g_s$ is set to $2.0$. 
Both equilibrium and out-of-equilibrium magnetization curves are presented in Fig.~\ref{fig:2_MB_J}. The out-of-equilibrium magnetization curve deviates markedly from the equilibrium one in all cases, reflecting the strength of the nonequilibrium effects. Over the magnetic field range of $0$--$1\,\textrm{T}$, the nonequilibrium effects are the weakest for the antisymmetric $\boldsymbol{J}^\textrm{dimer}_4$, moderate for the anisotropic $\boldsymbol{J}^\textrm{dimer}_2$, and the strongest for the isotropic $\boldsymbol{J}^\textrm{dimer}_1$ and anisotropic $\boldsymbol{J}^\textrm{dimer}_3$. The strength of the nonequilibrium effects may suggest the exchange coupling type, complementing the equilibrium magnetization. The magnetization for both $\boldsymbol{J}^\textrm{dimer}_1$ and $\boldsymbol{J}^\textrm{dimer}_3$ is almost zero over the whole magnetic field range. This is because the Zeeman term commutes with the exchange term at any time in both cases. 

\subsection{Three coupled spins of \texorpdfstring{$S=5/2$}{S=5/2}} 
\label{sec:Mn3}

An ongoing study by Lee \textit{et al.} investigates magnetic properties of a \ce{(CH6N3)2MnCl4} molecular crystal.~\cite{LeeSteppyMn3} The molecular unit is a Mn trimer with three local spins of $S=5/2$. Magnetization and EPR measurements were interpreted using the following spin Hamiltonian. 
\begin{equation}
\label{eq:HMn3}
\hat{H} = -2J(\hat{\mathbf{S}}_1 \cdot \hat{\mathbf{S}}_2 + \hat{\mathbf{S}}_2 \cdot \hat{\mathbf{S}}_3) + \sum_{i=1}^3 [ D \hat{S}_{i,z}^2 + E(\hat{S}_{i,x}^2 - \hat{S}_{i,y}^2)] + g_s \mu_B \mathbf{B} \cdot \hat{\mathbf{S}}
\end{equation}
In Eq.~\ref{eq:HMn3}, $J$ is the isotropic exchange coupling constant, and $D$ and $E$ are the second-order axial and rhombic ZFS parameters, respectively. $D$ and $E$ are related to the previously introduced ZFS parameters $B_k^q$ by $D=3B_2^0$ and $E=B_2^2$. $J$ was fitted to be $-2.42\,\textrm{cm}^{-1}$ using the measured magnetization, and the ZFS parameters were fitted to be $D=+0.167\, \textrm{cm}^{-1}$ and $E = +0.040\, \textrm{cm}^{-1}$ using the measured EPR spectra. Dynamical effects in the magnetization of this system will be reported in the same study by Lee \textit{et al.} Below, we use \ce{(CH6N3)2MnCl4} as an example to demonstrate the use of an effective basis.  

The dimension of the Hilbert space for the spin Hamiltonian in Eq.~\ref{eq:HMn3} is $(2\times 5/2+1)^3=216$. Under zero magnetic field, the eigenvalues of the spin Hamiltonian span over $130\, \textrm{cm}^{-1}$. The energy range is even bigger under finite magnetic fields. However, only the states within a few $\textrm{cm}^{-1}$ above the ground state are thermally relevant at low temperatures, say $0.6\, \textrm{K}$ at which experiments were performed. Therefore, it is not necessary to include all 216 eigenstates to calculate the magnetization. 
Tables~\ref{tab:basis1} and \ref{tab:basis2} show a set of 16 basis states and a set of 26 basis states, respectively, for setting up the effective Hamiltonian and modeling the magnetization. The states in both basis sets are chosen from the same pool of 216 states, which are the eigenstates of the isotropic exchange interaction plus a perturbative Zeeman term. The states in Table~\ref{tab:basis1} are selected using the procedure presented in Section~\ref{sec:basis}, while Table~\ref{tab:basis2} includes all states within $25\, \textrm{cm}^{-1}$ from the ground state in the pool. The state indices are \rev{1-based and} assigned \rev{in order of increasing energy} of all states in the pool. The candidate basis states are common eigenstates of the isotropic exchange coupling term\rev{, the $\hat{S}^2$ operator, and the $\hat{S}_z$ operator. The values of $S$ and $S_z$ in Tables~\ref{tab:basis1} and \ref{tab:basis2} are therefore quantum numbers. The accidental degeneracy between the $S=5/2$ and $S=1/2$ multiplets for the level at $24.20\,\textrm{cm}^{-1}$ is resolved by requiring the basis states to be eigenstates of $\hat{S}^2$ (see below Eq.~\ref{eq:X}).} 
\begin{table}[htb!]
\centering
\begin{tabular}{crrr}
\hline
Index & $E\, (\textrm{cm}^{-1}$) & $S\, (\hbar)$ & $S_z\, (\hbar)$ \\
\hline
  $  1$   &   $  0.00$   &   $2.5$   &   $-2.5$   \\
  $  2$   &   $  0.00$   &   $2.5$   &   $-1.5$   \\
  $  3$   &   $  0.00$   &   $2.5$   &   $-0.5$   \\
  $  4$   &   $  0.00$   &   $2.5$   &   $ 0.5$   \\
  $  5$   &   $  0.00$   &   $2.5$   &   $ 1.5$   \\
  \rev{$  6$}   &   $  0.00$   &   $2.5$   &   $ 2.5$   \\
  \rev{$ 11$}   &   $ 16.94$   &   $3.5$   &   $-3.5$   \\
  \rev{$ 18$}   &   $ 16.94$   &   $3.5$   &   $ 3.5$   \\
  \rev{$ 31$}   &   $ 38.72$   &   $4.5$   &   $-4.5$   \\
  \rev{$ 42$}   &   $ 38.72$   &   $4.5$   &   $ 4.5$   \\
  \rev{$ 89$}   &   $ 65.34$   &   $5.5$   &   $-5.5$   \\
  \rev{$100$}   &   $ 65.34$   &   $5.5$   &   $ 5.5$   \\
  \rev{$151$}   &   $ 96.80$   &   $6.5$   &   $-6.5$   \\
  \rev{$174$}   &   $ 96.80$   &   $6.5$   &   $ 6.5$   \\
  \rev{$201$}   &   $133.10$   &   $7.5$   &   $-7.5$   \\
  $216$   &   $133.10$   &   $7.5$   &   $ 7.5$   \\
\hline
\end{tabular}
\caption{\label{tab:basis1}The energy, total spin, and $z$-projection of spin of the 16 basis states selected for setting up an effective Hamiltonian.}
\end{table}
\begin{table}[htb!]
\centering
\begin{tabular}{crrr}
\hline
Index & $E\, (\textrm{cm}^{-1}$) & $S\, (\hbar)$ & $S_z\, (\hbar)$ \\
\hline
  $  1$   &   $  0.00$   &   $2.5$   &   $-2.5$   \\
  $  2$   &   $  0.00$   &   $2.5$   &   $-1.5$   \\
  $  3$   &   $  0.00$   &   $2.5$   &   $-0.5$   \\
  $  4$   &   $  0.00$   &   $2.5$   &   $ 0.5$   \\
  $  5$   &   $  0.00$   &   $2.5$   &   $ 1.5$   \\
  $  6$   &   $  0.00$   &   $2.5$   &   $ 2.5$   \\
  $  7$   &   $ 12.10$   &   $1.5$   &   $-1.5$   \\
  $  8$   &   $ 12.10$   &   $1.5$   &   $-0.5$   \\
  $  9$   &   $ 12.10$   &   $1.5$   &   $ 0.5$   \\
  $ 10$   &   $ 12.10$   &   $1.5$   &   $ 1.5$   \\
  $ 11$   &   $ 16.94$   &   $3.5$   &   $-3.5$   \\
  $ 12$   &   $ 16.94$   &   $3.5$   &   $-2.5$   \\
  $ 13$   &   $ 16.94$   &   $3.5$   &   $-1.5$   \\
  $ 14$   &   $ 16.94$   &   $3.5$   &   $-0.5$   \\
  $ 15$   &   $ 16.94$   &   $3.5$   &   $ 0.5$   \\
  $ 16$   &   $ 16.94$   &   $3.5$   &   $ 1.5$   \\
  $ 17$   &   $ 16.94$   &   $3.5$   &   $ 2.5$   \\
  $ 18$   &   $ 16.94$   &   $3.5$   &   $ 3.5$   \\
  $ 19$   &   $ 24.20$   &   $2.5$   &   $-2.5$   \\
  $ 20$   &   $ 24.20$   &   $2.5$   &   $-1.5$   \\
  $ 21$   &   $ 24.20$   &   \rev{$0.5$}   &   $-0.5$   \\
  $ 22$   &   $ 24.20$   &   \rev{$2.5$}   &   $-0.5$   \\
  $ 23$   &   $ 24.20$   &   \rev{$0.5$}   &   $ 0.5$   \\
  $ 24$   &   $ 24.20$   &   \rev{$2.5$}   &   $ 0.5$   \\
  $ 25$   &   $ 24.20$   &   $2.5$   &   $ 1.5$   \\
  $ 26$   &   $ 24.20$   &   $2.5$   &   $ 2.5$   \\
\hline
\end{tabular}
\caption{\label{tab:basis2}The energy, total spin, and $z$-projection of the spin of the 26 basis states selected for setting up an effective Hamiltonian.}
\end{table}

\begin{figure}[htb!]
\centering
\includegraphics[width=0.6\columnwidth]{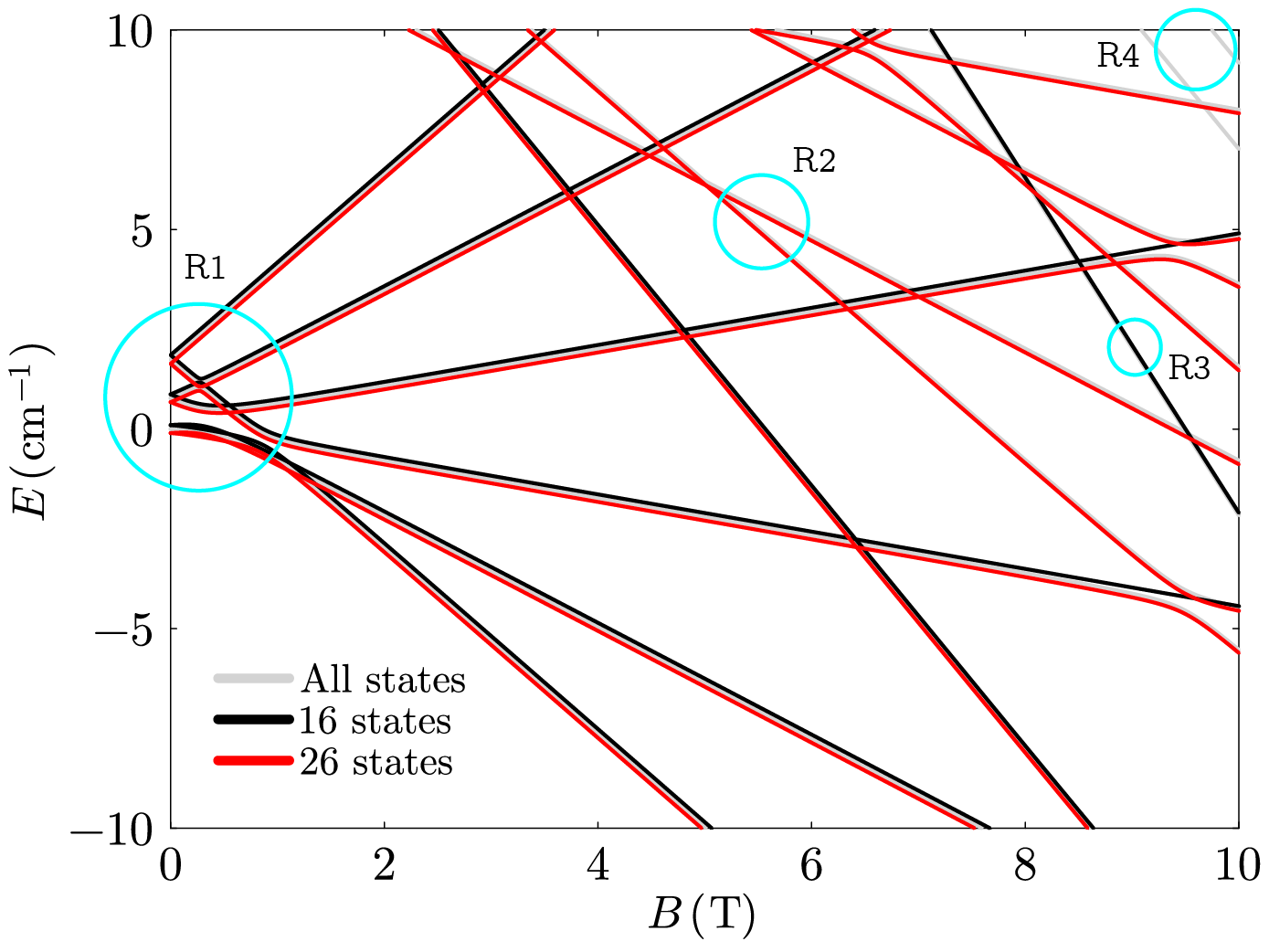}
\caption{\label{fig:3_Zeeman}Zeeman energy diagram for \ce{(CH6N3)2MnCl4} using the full spin Hamiltonian and two effective Hamiltonians. The energies are offset slightly to show the overlapping lines. Gray lines underlie every black or red line. Four regions in the plot are highlighted by cyan circles.}
\end{figure}

\begin{figure}[htb!]
\centering
\includegraphics[width=0.6\columnwidth]{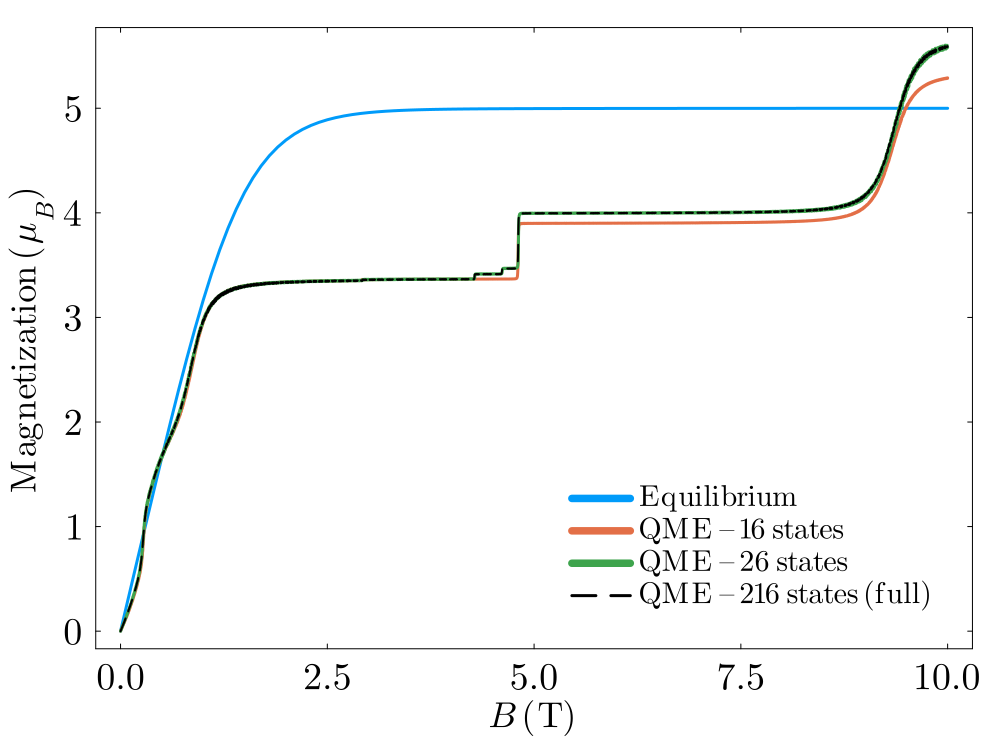}
\caption{\label{fig:3_MB_basis}Magnetization versus magnetic field for \ce{(CH6N3)2MnCl4} using \rev{three} different sets of basis states. Parameters used in calculations: $T=0.6~\textrm{K}$, $I_0 = 1\times 10^{-14}~{\textrm{ps}^{3}\cdot \textrm{rad}^{-2}}$, sweep rate $=50~\textrm{T}/\textrm{ms}$, and time step $=\rev{1\times 10^{3}}~\mathrm{ps}$. The equilibrium magnetization is also plotted for comparison. \rev{The 16- and 26-state curves are calculated with the Pad\'{e} propagator and the full-space curve with the Strang propagator.}}
\end{figure}

Figure~\ref{fig:3_Zeeman} shows the energy levels versus magnetic field as calculated using the full spin Hamiltonian of dimensions $216\times 216$, the first effective Hamiltonian of dimensions $16\times 16$, and the second effective Hamiltonian of dimensions $26\times 26$. 
Four different regions of the energy level diagram are highlighted by cyan circles and marked by R1, R2, R3, and R4. In the region R1 where both the magnetic field and the energy are low, all six energy levels are well reproduced by both effective Hamiltonians. In the whole magnetic field range, the lowest three energy levels are well reproduced. This provides a basis for faithfully calculating the magnetization dynamics using the effective Hamiltonians. 
However, certain energy levels are missing from the solutions of the effective Hamiltonians. For example, the first effective Hamiltonian does not reproduce the energy levels in the region R2, the second effective Hamiltonian fails to reproduce an energy level in the region R3, and both effective Hamiltonians do not yield the energy levels in the region R4. It limits the application of the effective Hamiltonians at high temperatures, at which the missing energy levels could be thermally occupied according to the Boltzmann distribution. 
Fig.~\ref{fig:3_MB_basis} shows the calculated magnetization versus magnetic field at $0.6\, \textrm{K}$ using both effective Hamiltonians. The two magnetization curves are close to each other before $9\,\mathrm{T}$, both exhibiting lower magnetization plateaus than the magnetization plateau under equilibrium and a sizable sudden increase in the magnetization at $\sim 5\,\mathrm{T}$. However, they exhibit different minor sudden increases from $2.5\,\mathrm{T}$ to $5\,\mathrm{T}$ and deviate from each other significantly near $10\,\mathrm{T}$. The second effective Hamiltonian of dimensions $26\times 26$ should be more reliable in this circumstance since it includes more thermally relevant states. \rev{This is confirmed by comparison with the full-space solution, which was obtained with the approximate Strang propagator (see \ref{sec:propagators}). Solving the quantum master equation in the full space is beyond the reach of the exact propagator, since the Liouvillian dimension is as high as $2\times 216^2 = 93312$.} At even higher magnetic fields, the results from the first effective Hamiltonian will be more reliable despite its smaller dimensions, since it can describe the states with $|m_s| > 7/2$ which are beyond the description of the second effective Hamiltonian. The states with a larger $|m_s|$ are energetically preferred under higher magnetic fields. 

\rev{We recommend the following procedure for choosing an effective basis. First, include all pool states whose energies are within a cutoff $E_\text{cut}$ of the instantaneous ground state at some field of the sweep. $E_\text{cut}$ should be several times larger than both $k_BT$ and the excitation energies of the states that become populated during the sweep. Second, compare the Zeeman diagram and the equilibrium magnetization with those of the full Hamiltonian. Third, enlarge the basis until the nonequilibrium magnetization is converged to the desired tolerance. If the full Hilbert space is small enough, the Strang propagator provides a direct reference.}

\section{Conclusions} 

In summary, we have presented \textit{qdmag}, a Python package for simulating nonequilibrium magnetization dynamics of magnetic molecules under a time-varying external magnetic field in the presence of spin-phonon coupling.
The package implements the generalized Lindblad quantum master equation of Saito \textit{et al.}, supports a wide range of magnetic exchange interactions and extended Stevens operators up to 12\textsuperscript{th} order, and accommodates four types of magnetic field profiles including pulsed and sinusoidal fields. 
To make the calculations computationally feasible for large spin systems, \textit{qdmag} employs a staircase approximation and an effective Hamiltonian scheme, enabling long-time magnetization dynamics over timescales of up to a few milliseconds. 
We demonstrated the capabilities of the package through three case studies: a mononuclear \ce{Ho^{3+}} complex with $J=8$, two spins of $S=1/2$ coupled by four types of exchange coupling, and a Mn trimer with $S=5/2$ local spins. The case studies illustrate the powder-averaging functionality, the construction and validation of effective Hamiltonians, and the numerical stability and efficiency of the staircase approximation over the conventional Runge--Kutta method. 
\textit{qdmag} bridges the gap between theoretical formalism and practical numerical
simulation of spin dynamics, and is expected to be a valuable tool for interpreting pulsed-field magnetometry experiments and understanding spin-phonon-driven relaxation in magnetic molecules.

\appendix

\section{Conditions of \texorpdfstring{$\rho$}{ρ}} 
\label{sec:conditions}

It is known that the reduced density matrix satisfies the conditions given below.~\cite{Blum2012} 
\begin{enumerate}
\item Normalization: $\mathrm{Tr}(\rho) = 1$ or $\sum_i \lambda_i = 1$. 
\item Hermiticity: $\rho^\dagger = \rho$.
\item Nonnegativity: $\rho$ is positive semi-definite, i.e. $\lambda_i \ge 0$ $\forall$ $i$. 
\item Cauchy-Schwarz inequality: $|\rho|_{ij}^2 \le \rho_{ii} \rho_{jj} $ $\forall$ $i$ and $j$.
\end{enumerate}
In the above list, $\lambda_i$ are the eigenvalues of the reduced density matrix $\rho$. The last condition follows from the more general Cauchy-Schwarz inequality in quantum mechanics. 

\section{Comparison with the Runge-Kutta method} 
\label{sec:RK4}

\begin{figure}[htb!]
\centering
\includegraphics[width=0.6\columnwidth]{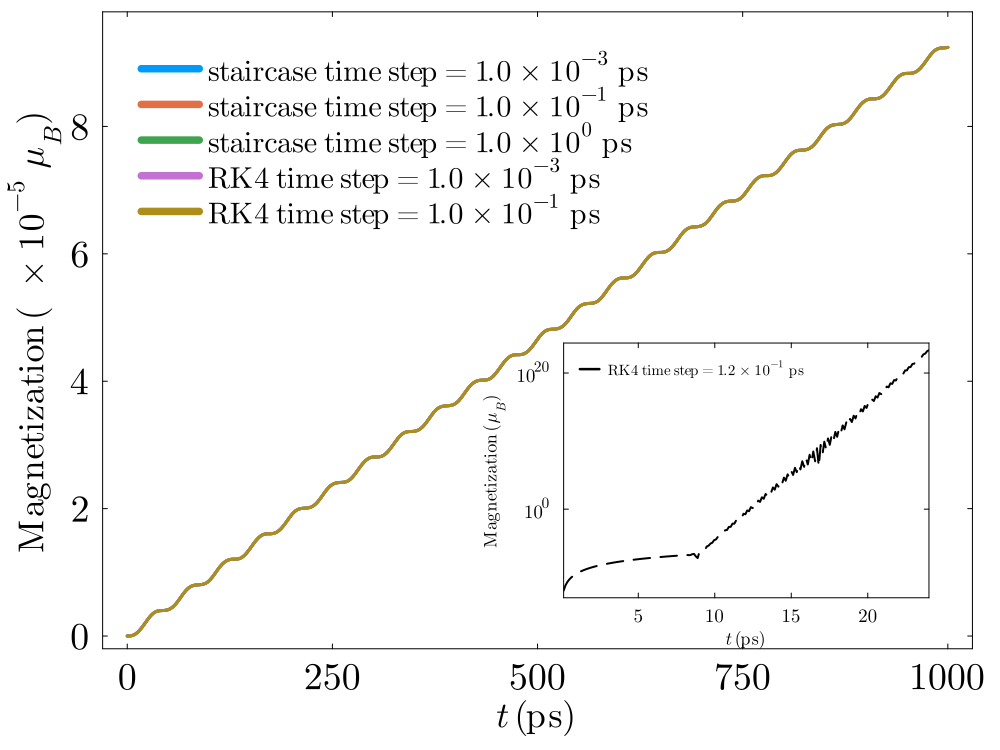}
\caption{\label{fig:3_Mt_RK4}Magnetization versus time for \ce{(CH6N3)2MnCl4} using both the fourth-order Runge-Kutta (RK4) method and the staircase approximation. All five curves in the plot collapse onto the same line. Inset: a divergent magnetization curve obtained by RK4. Parameters used in calculations: $T=0.6~\textrm{K}$, $I_0 = 1\times 10^{-14}~{\textrm{ps}^{3}\cdot \textrm{rad}^{-2}}$, and sweep rate $=50~\textrm{T}/\textrm{ms}$.}
\end{figure}

The quantum master equation in the Liouville form on the basis of the 16 states was solved for \ce{(CH6N3)2MnCl4} using both the fourth-order Runge-Kutta method (RK4) and the staircase approximation. As shown in Fig.~\ref{fig:3_Mt_RK4}, the two methods yield consistent results using a time step of $0.1\, \textrm{ps}$ or a smaller time step. If the time step is $0.12\, \textrm{ps}$, as shown in the inset of Fig.~\ref{fig:3_Mt_RK4}, the magnetization calculated using the RK4 method will diverge. In contrast, the staircase approximation is still reliable with a time step of $1\, \textrm{ps}$. Given that each time step takes more than 20 seconds using 16 AMD EPYC 75F3 cores, it is computationally infeasible to evolve the density matrix from zero to $1\,\textrm{ms}$ with the RK4 method. 

\section{Comparison of the exponential propagators} 
\label{sec:propagators}

\textit{qdmag} offers the following three methods to evaluate the action of the matrix exponential $\textrm{exp}(\mathcal{L} \Delta t)$ on the vectorized density matrix. 
\begin{enumerate}
\item \textbf{Pad\'{e}.} The matrix exponential is formed in full with the scaling-and-squaring Pad\'{e} approximant~\cite{Higham2005Pade, AlMohy2009Pade}, and is then applied to $\rho$. 
\item \textbf{Taylor.} The action of the exponential is evaluated with a truncated Taylor series without ever forming $\textrm{exp}(\mathcal{L} \Delta t)$~\cite{AlMohy2011Action}.
\item \textbf{Krylov.} The action is evaluated by projecting onto a Krylov subspace of lower dimension than the Liouville space, generated via the Arnoldi process~\cite{Saad1992Krylov, Sidje1998Expokit}.
\end{enumerate}
The Taylor and Krylov methods are additionally available in a sparse variant, in which $\mathcal{L}$ is built and held in the compressed sparse row format. 
\rev{As will be shown below, the} large time steps that make the staircase approximation efficient are precisely the regime in which the Pad\'{e} method outperforms the other two methods.
The Taylor and Krylov methods become preferable for short time steps.

\begin{figure}[htb!]
\centering
\includegraphics[width=\linewidth]{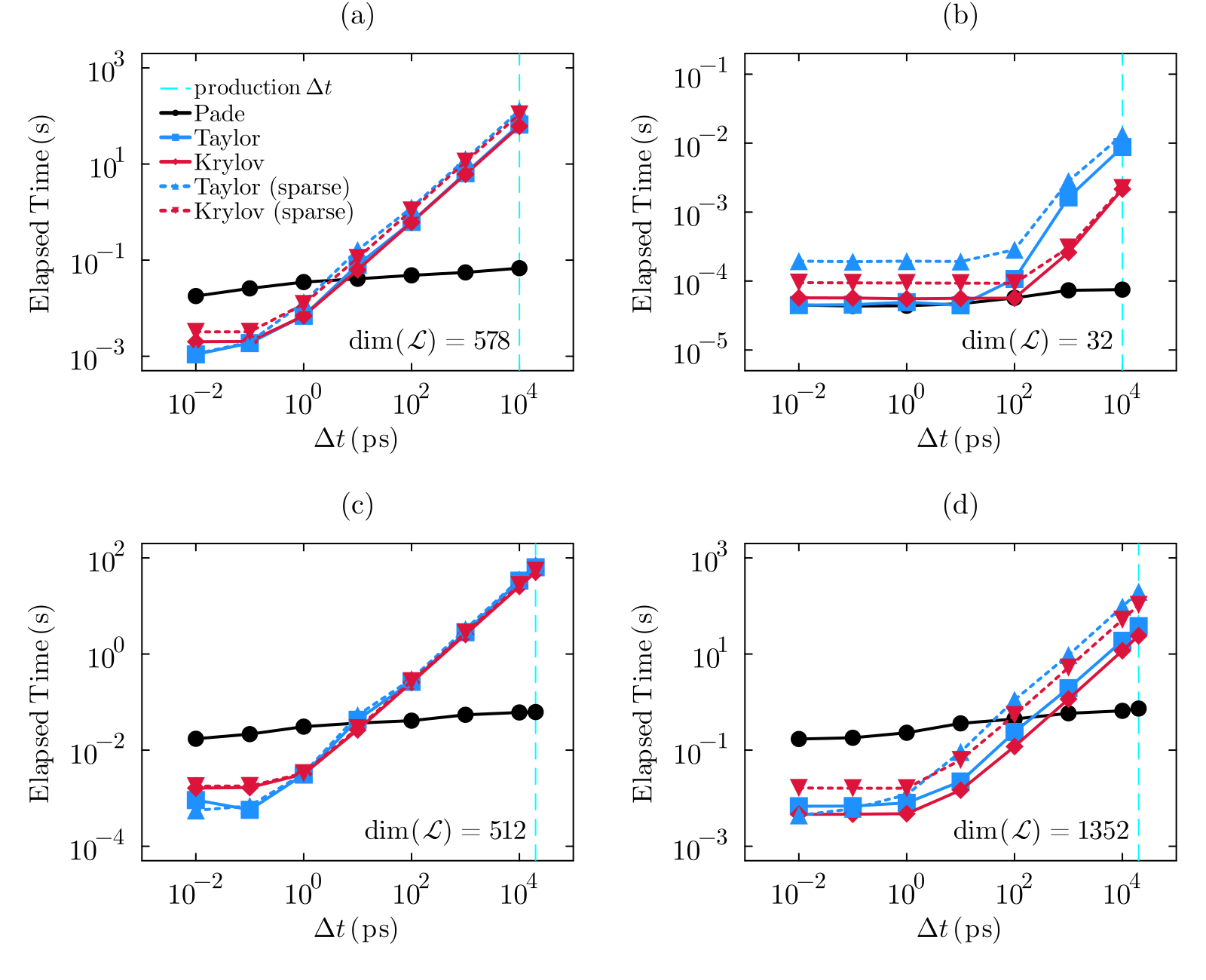}
\caption{\label{fig:cost_vs_deltat}
Wall time of a single application of $\textrm{exp}(\mathcal{L} \Delta t)$ to the vectorized density matrix versus the time step, for the five numerical methods and the following four systems: (a) \ce{Ho(pzdo)4}, (b) the $S=1/2$ dimer, and \ce{(CH6N3)2MnCl4} in the (c) 16-state and (d) 26-state effective basis. The dimension of the Liouville space is annotated in each panel, and the dashed vertical line marks the time step used in the main text.}
\end{figure} 
Figure~\ref{fig:cost_vs_deltat} compares the performance of the three methods of calculating $\exp{(\mathcal{L}\Delta t)\rho}$ and their two variants using the four aforementioned effective Hamiltonians. 
Each point in the figure is the median of three evaluations when the runtime exceeds $50~\textrm{ms}$, a single evaluation when the runtime exceeds $20~\textrm{s}$, and as many calls as fit into $\sim 0.5~\textrm{s}$ when the runtime is below $50~\textrm{ms}$. 
As the time step increases from $1\times 10^{-2}\,\textrm{ps}$ to $1\times 10^{4}\,\textrm{ps}$, the cost of the Pad\'{e} propagator is nearly flat, varying by no more than a factor of $4.3$ across the whole sweep in any of the four panels, whereas the cost of the Taylor and the Krylov propagators grow by 2--5 orders of magnitude depending on the system. 
Below $\Delta t \approx 1~\textrm{ps}$ the latter two flatten onto a floor set by the fixed cost of a single substep, where they are most advantageous. 
At the time step adopted in the main text, the Pad\'{e} propagator is the fastest of the five for all four systems, by factors of $899$, $29$, $815$, and $33$ over the fastest alternative, which is the Krylov propagator in every case. 
The sparse variants lie above their dense counterparts at the production time step in all four panels due to the modest sparsity of $\mathcal{L}$. 
All these methods agree on the propagated density matrix to within $1\times10^{-11}$ for the results presented above. 
The timings were obtained on a single core of an Apple M3 processor.

\rev{All three methods above act on the vectorized density matrix with the Liouvillian $\mathcal{L}$ of dimension $2d^2$, where $d$ is the dimension of the (effective) Hilbert space, and the Pad\'{e} method stores it densely. This limits them to systems with $d$ of at most about $100$. For larger systems, \textit{qdmag} offers an approximate propagator which works on $\rho$ as a $d\times d$ matrix and never forms $\mathcal{L}$. The approximate propagator relies on the symmetric (Strang) splitting~\cite{Strang1968} of $\mathcal{L}$ as implemented in Spinach~\cite{Spinach2026PulsedField}. The splitting is exact if the coherent and dissipative parts commute. Otherwise, its error over a fixed time interval is of second order. The $216$-state solution of \ce{(CH6N3)2MnCl4} is feasible, given that a stair takes $25$--$45~\textrm{ms}$ with the Strang propagator on two cores. The Strang and Pad\'{e} propagators agree to within $10^{-6}\,\mu_B$ in magnetization from $0$ to $10~\textrm{T}$, at time steps of $2\times10^{4}$ and $10^{3}~\textrm{ps}$ in the 16- and 26-state model spaces at $0.6~\textrm{K}$ and $50~\textrm{T}/\textrm{ms}$.}

\section{Further Analysis for \ce{Ho(pzdo)4}} 
\label{sec:HoAna}

\begin{figure}[htb!]
\centering
\includegraphics[width=0.75\columnwidth]{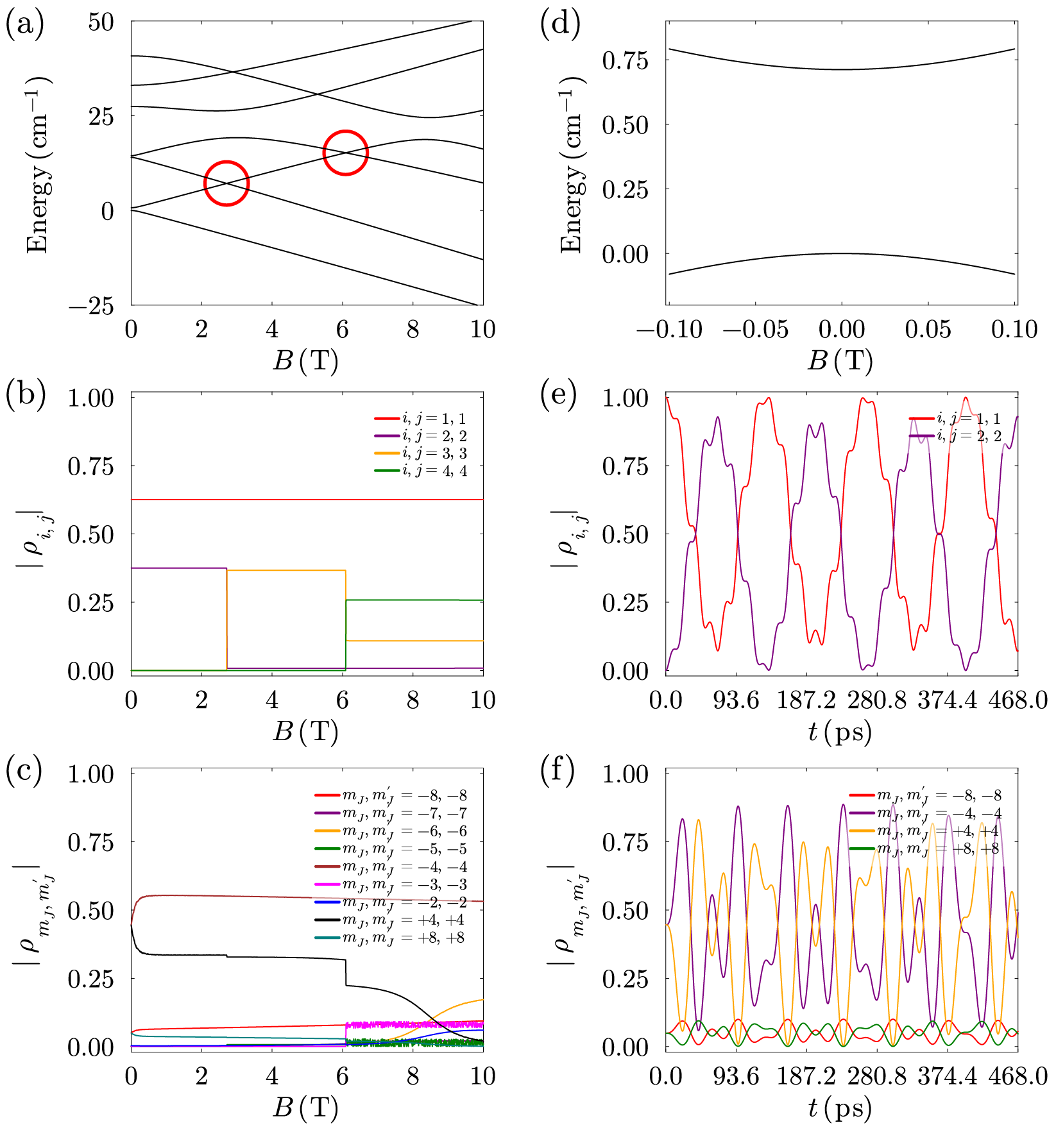}
\caption{\label{fig:1_zeeman_rho}(a) Zeeman energy diagram for \ce{Ho(pzdo)4}. Two \rev{avoided} crossings are highlighted by red circles. (b) Absolute values of the relevant matrix elements of the density matrix $\rho$ in the basis of the instantaneous eigenstates of the Hamiltonian and (c) the $\hat{J}_z$ operator for the complex under the linear magnetic field profile shown in Fig.~\ref{fig:1_MB_pulse}a in the main text. (d) A zoomed-in version of the Zeeman energy diagram to show the lowest two energy levels. (e) and (f) are similar to (b) and (c) but for the oscillating magnetic field in Fig.~\ref{fig:1_MB_pulse}d in the main text.
}
\end{figure}

The Zeeman energy diagram for \ce{Ho(pzdo)4} under a magnetic field along the main magnetic $z$-axis up to 10 T is shown in Fig.~\ref{fig:1_zeeman_rho}a. The red circles highlight two \rev{avoided} crossings, one between the first and second excited states and one between the second and third excited states. Both \rev{avoided} crossings give rise to dramatic changes in the occupation numbers of the relevant energy levels, see Fig.~\ref{fig:1_zeeman_rho}b. As shown in Fig.~\ref{fig:1_zeeman_rho}c, the occupation of the eigenstates of the $\hat{J}_z$ operator also changes dramatically, enhancing the imbalance between $+|M_J|$ and $-|M_J|$ states. This explains the sudden jumps in the magnetization curve shown in Fig.~\ref{fig:1_MB_pulse}a. Although the occupation numbers of the ground and first excited states remain nearly constant before the first \rev{avoided} crossing at $B = 2.7\,\mathrm{T}$, the occupation of $M_J = -4 (+4)$ increases (decreases) before $B = 0.5\,\mathrm{T}$, leading to a gradual increase of the magnetization from zero to $1.3\,\mu_B$. Between $0.5\,\mathrm{T}$ and $2.7\,\mathrm{T}$, the reduced density matrix remains nearly constant, and a magnetization plateau is therefore formed over this field range. The magnetization plateau between the two \rev{avoided} crossing points can be understood in a similar manner. 

Fig.~\ref{fig:1_zeeman_rho}d shows the lowest two energy levels versus the magnetic field from $-0.1$ T to $0.1$ T. The sinusoidal magnetic field $B=B_0 \sin(\omega t)$ in Fig.~\ref{fig:1_MB_pulse}d drives the system to oscillate between these two energy levels periodically, see Fig.~\ref{fig:1_zeeman_rho}e. The period of oscillation in the occupation of these two levels is 136.8 ps, which is longer than but not a multiple of the period of the applied magnetic field (46.8 ps). As a result, the occupation of the eigenstates of the $\hat{J}_z$ operator does not exhibit periodicity within 468.0 ps, see Fig.~\ref{fig:1_zeeman_rho}f. Accordingly, the magnetization curve in Fig.~\ref{fig:1_MB_pulse}d does not exhibit periodicity. It is noteworthy that the applied magnetic field alters the composition of the two levels in terms of the eigenstates of the $\hat{J}_z$ operator; $\rho_{m_J,m_J}$ would oscillate with the same period as the applied magnetic field if the occupation of the two levels remained constant. 

The two energy levels in Fig.~\ref{fig:1_zeeman_rho}d can be described by the following effective two-level model
\begin{equation}
    \hat{H}_\textrm{eff}(\mathbf{B}) = \frac{\Delta}{2}\,\sigma_z + \tfrac12\,g_\parallel\,\mu_B B_z\,\sigma_x,
\end{equation}
where $\Delta$ is the clock transition gap, $\sigma_x$ and $\sigma_z$ are Pauli matrices, and $g_\parallel$ is an effective $g$-factor. $g_\parallel$ is calculated to be 10.785 using the method of Chibotaru and Ungur~\cite{Chibotaru2012SINGLEANI}.
Such a two-level model yields Rabi oscillations with a frequency $f_\textrm{Rabi}=(\mu_B/h) g_\parallel B_z/2$ under the rotating wave approximation (RWA). The RWA Rabi period at 0.1 T is 132.5 ps, which is close to the \textit{qdmag} result of 136.8 ps. The Rabi period by \textit{qdmag} increases by a factor of 9.7 instead of exactly 10 times when $B_0$ decreases from 0.1 T to 0.01 T. The discrepancy between the \textit{qdmag} solution and the analytic RWA solution is likely due to the fact that higher energy levels are missing from the effective two-level model. A change of basis from the instantaneous eigenstates of the Hamiltonian to the eigenstates of the initial Hamiltonian does not resolve the discrepancy. 


\section{Convergence with respect to the time step} 
\label{sec:convergence}

\begin{figure}[htb!]
\centering
\includegraphics[width=0.6\columnwidth]{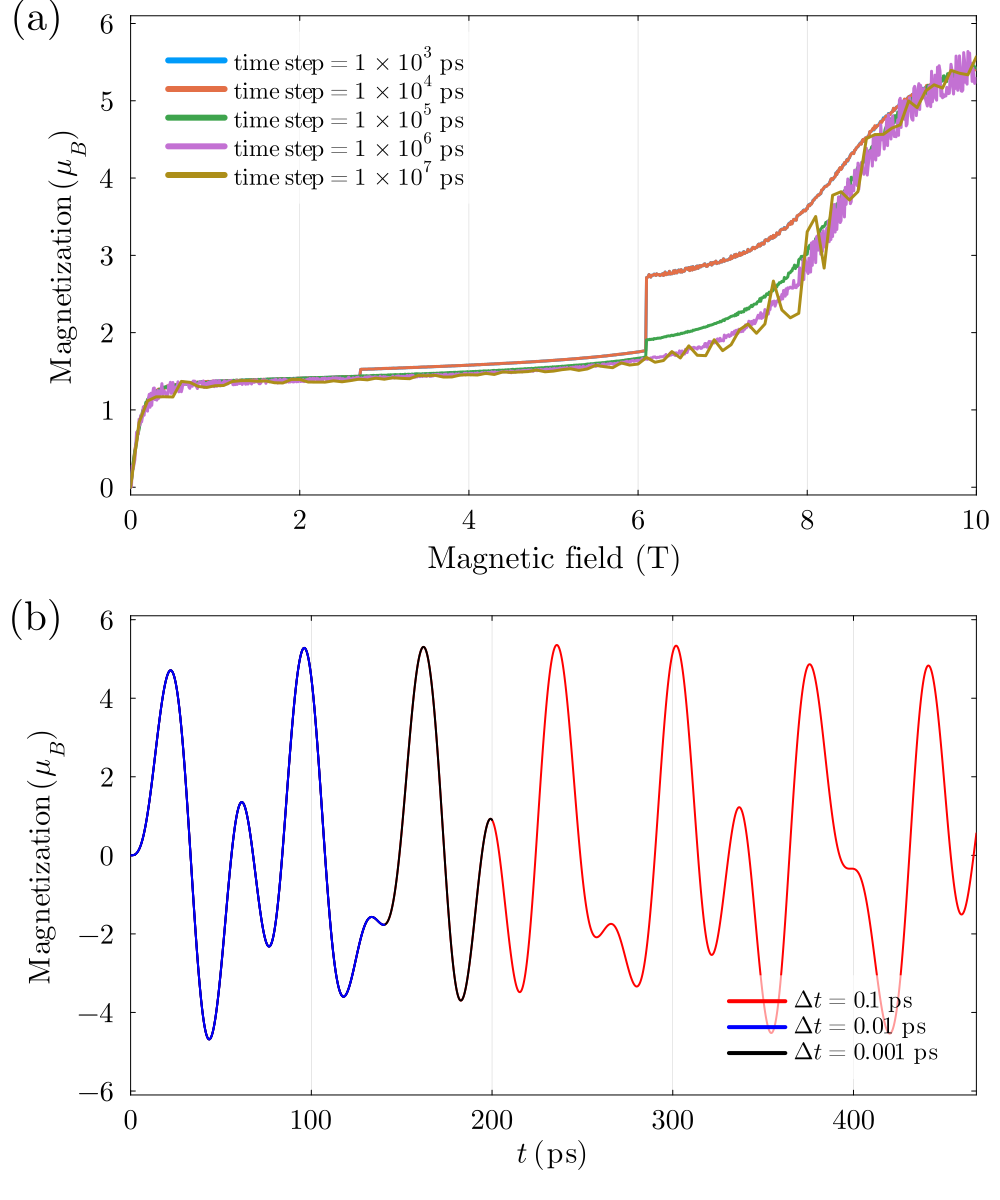}
\caption{\label{fig:1_MB_step}Convergence of magnetization against time step in the staircase approximation. Magnetization of \ce{Ho(pzdo)4} under (a) a linear magnetic field profile and (b) an oscillating magnetic field. The magnetic field is applied along the pseudo four-fold rotational symmetry axis of the complex. \rev{The sweep rate in (a) is $10~\textrm{T}/\textrm{ms}$, so the time steps of $10^{3}$ to $10^{7}~\textrm{ps}$ correspond to field steps of $10^{-5}$ to $10^{-1}~\textrm{T}$ per stair.}}
\end{figure}

Figs.~\ref{fig:1_MB_step}a and \ref{fig:1_MB_step}b show the calculated magnetization of \ce{Ho(pzdo)4} under the linear magnetic field profile in Fig.~\ref{fig:1_MB_pulse}a\rev{, with a sweep rate of $10~\textrm{T}/\textrm{ms}$,} and the sinusoidal magnetic field in Fig.~\ref{fig:1_MB_pulse}d, respectively, using different time steps in the staircase approximation. For the linear magnetic field profile, a time step of \rev{$1\times 10^{4}~\textrm{ps}$} yields converged magnetization. The magnetic field increases by $0.0001~\textrm{T}$ in each time step. For the sinusoidal magnetic field, a time step of 0.1 ps is shown to be sufficiently small to yield a reliable magnetization curve. It is noteworthy that the choice of time step depends on the Hamiltonian. Additional test calculations show that a much longer time step can be used for an isotropic Heisenberg Hamiltonian. 

\rev{For \ce{Ho(pzdo)4} under the linear magnetic field sweep, the time step is limited by the avoided level crossings. Near an avoided crossing at $B_0$, the gap between the two levels is $[\Delta^2 + \kappa^2 (B-B_0)^2]^{1/2}$, where $\Delta$ is the minimal gap and $\kappa$ is the slope of the gap far from the avoided crossing. The eigenstates change within a magnetic-field window of width $\delta B = \Delta/\kappa$. The probability of a diabatic passage is given by the Landau--Zener formula $P = \exp[-\pi\Delta^2/(2\hbar\, \kappa\, \dot{B})]$ where $\dot{B}$ is the sweep rate. The avoided crossings at $2.71$ and $6.09~\textrm{T}$ in Fig.~\ref{fig:1_zeeman_rho}a have $\Delta = 6.1\times 10^{-5}$ and $2.2\times 10^{-4}~\textrm{cm}^{-1}$ and $\delta B = 1.2\times 10^{-5}$ and $5.4\times 10^{-5}~\textrm{T}$, respectively. At a sweep rate of $10~\textrm{T}/\textrm{ms}$, the Landau--Zener formula gives $P=0.979$ and $0.703$ for those two avoided crossings. We swept the field across each crossing within $\pm 0.01~\textrm{T}$, starting from the upper level. A field step of $10^{-4}~\textrm{T}$ per stair yields $P=0.978$ and $0.704$. A field step of $10^{-3}~\textrm{T}$ instead yields $P=0.999$ and $0.989$, i.e., an artificially diabatic passage. At a sweep rate of $50~\textrm{T}/\textrm{ms}$, the formula gives $P=0.932$ for the second avoided crossing, which a field step of $10^{-4}~\textrm{T}$ reproduces ($0.933$) and a field step of $10^{-3}~\textrm{T}$ does not ($0.988$). We therefore recommend that the field step per stair be no larger than about twice the width $\delta B$ of the narrowest avoided crossing between the populated levels, which can be estimated in advance from the Zeeman diagram.} 

\section{Convergence with respect to the Lebedev order} 
\label{sec:powderconv}

Table~\ref{tab:powderconv} lists the powder-averaged equilibrium magnetization of \ce{Ho(pzdo)4} at 2, 10, and 50 T obtained with Lebedev quadratures of order 5 to 47. From order 21 onwards, all three magnetic fields are converged to within $0.006\, \mu_B$ of the order-47 result, and the residual change between orders 31 and 47 is below $0.003\, \mu_B$. An order of 21, corresponding to 170 orientations, is therefore adopted in this work as a compromise between accuracy and computational cost.

\begin{table}[htb!]
\centering
\begin{tabular}{crrrr}
\hline
\multirow{2}{*}{Order} & \multirow{2}{*}{Points} & \multicolumn{3}{c}{$\overline{M}\, (\mu_B)$} \\
\cline{3-5}
 &  & $2$ T & $10$ T & $50$ T \\
\hline
  $ 5$   &   $ 14$   &   $4.783$   &   $7.646$   &   $9.083$   \\
  $ 9$   &   $ 38$   &   $4.763$   &   $7.653$   &   $8.884$   \\
  $15$   &   $ 86$   &   $4.786$   &   $7.706$   &   $8.967$   \\
  $21$   &   $170$   &   $4.787$   &   $7.719$   &   $8.976$   \\
  $31$   &   $350$   &   $4.787$   &   $7.722$   &   $8.980$   \\
  $47$   &   $770$   &   $4.787$   &   $7.724$   &   $8.982$   \\
\hline
\end{tabular}
\caption{\label{tab:powderconv}Convergence of the powder-averaged magnetization $\overline{M}$ of \ce{Ho(pzdo)4} in thermal equilibrium at $T=2\, \textrm{K}$ with respect to the order of the Lebedev quadrature used in Eq.~\ref{eq:powder}. The number of quadrature points associated with each order is also given.}
\end{table}

\rev{
\section{Installation and the input file}
\label{sec:input}

\textit{qdmag} is available on the Python Package Index (PyPI) and can be installed with \texttt{pip install qdmag}, which also installs its dependencies (NumPy, SciPy $\geq 1.15$, PyYAML, h5py, and pandas). The version described in this work is 1.3.0. The input files of the case studies and the test suite are not part of the package and are distributed with the repository, from whose root directory \textit{qdmag} can also be installed with \texttt{pip install .}.
A calculation is run in a working directory that contains the input file \texttt{input.yaml}, e.g., \texttt{python -m qdmag.tools.tool\_staircase} for the time evolution. The input specifications are described on the documentation webpage at \url{https://shuanglongliu.github.io/qdmag/input.html}. Energies and coupling constants are in $\textrm{cm}^{-1}$, magnetic fields in T, times in ps, and temperatures in K. Each \texttt{reference\_frame} is a $3\times 3$ matrix, given row by row, whose rows are the local $x$, $y$, and $z$ axes expressed in the lab frame. Site indices start from 1, and the indices of the pool states in the \texttt{states} key start from 0.

}

\section*{Acknowledgments}

\rev{We thank Ilya Kuprov for bringing the pulsed-field functionality of Spinach to our attention.} This work is supported by the Center for Molecular Magnetic Quantum Materials (M2QM), an Energy Frontier Research Center (EFRC) funded by the U.S. Department of Energy (DOE), Office of Science, Basic Energy Sciences under Award DE-SC0019330. The calculations employed resources of the University of Florida Research Computing as well as the National Energy Research Scientific Computing Center (NERSC), a Department of Energy User Facility using NERSC award BES-ERCAP0032450. 

\bibliography{references}

\begin{thebibliography}{10}
\expandafter\ifx\csname url\endcsname\relax
  \def\url#1{\texttt{#1}}\fi
\expandafter\ifx\csname urlprefix\endcsname\relax\def\urlprefix{URL }\fi
\expandafter\ifx\csname href\endcsname\relax
  \def\href#1#2{#2} \def\path#1{#1}\fi

\bibitem{Shu2023Radicals}
C.~Shu, Z.~Yang, A.~Rajca, From stable radicals to thermally robust high-spin
  diradicals and triradicals, Chemical Reviews 123~(20) (2023) 11954--12003.
\newblock \href {https://doi.org/10.1021/acs.chemrev.3c00406}
  {\path{doi:10.1021/acs.chemrev.3c00406}}.

\bibitem{mabbs1973magnetism}
F.~Mabbs, D.~Machin, Magnetism and Transition Metal Complexes, Chapman and
  Hall, 1973.

\bibitem{Benelli2015MagMol}
C.~Benelli, D.~Gatteschi, Introduction to Molecular Magnetism, Wiley-VCH Verlag
  GmbH \& Co. KGaA, 2015.
\newblock \href {https://doi.org/10.1002/9783527690541}
  {\path{doi:10.1002/9783527690541}}.

\bibitem{Woodruff2013-LnMM}
D.~N. Woodruff, R.~E.~P. Winpenny, R.~A. Layfield, Lanthanide single-molecule
  magnets, Chemical Reviews 113~(7) (2013) 5110--5148.
\newblock \href {https://doi.org/10.1021/cr400018q}
  {\path{doi:10.1021/cr400018q}}.

\bibitem{Hu2024-LnMM}
Z.~Hu, S.~Yang, Endohedral metallofullerene molecular nanomagnets, Chemical
  Society Reviews 53~(6) (2024) 2863--2897.
\newblock \href {https://doi.org/10.1039/d3cs00991b}
  {\path{doi:10.1039/d3cs00991b}}.

\bibitem{Li2024-LnMM}
J.~Li, Y.~Yang, Q.~Yu, G.~Su, W.~Liu, Development and prospects in the
  structure and performance of lanthanide single-molecule magnets, The Journal
  of Physical Chemistry C 128~(12) (2024) 4882--4890.
\newblock \href {https://doi.org/10.1021/acs.jpcc.4c00506}
  {\path{doi:10.1021/acs.jpcc.4c00506}}.

\bibitem{Shiddiq2016-HoW10}
M.~Shiddiq, D.~Komijani, Y.~Duan, A.~Gaita-Ariño, E.~Coronado, S.~Hill,
  Enhancing coherence in molecular spin qubits via atomic clock transitions,
  Nature 531~(7594) (2016) 348--351.
\newblock \href {https://doi.org/10.1038/nature16984}
  {\path{doi:10.1038/nature16984}}.

\bibitem{Stewart2024-HoLF}
R.~Stewart, A.~B. Canaj, S.~Liu, E.~Regincós~Martí, A.~Celmina, G.~Nichol,
  H.-P. Cheng, M.~Murrie, S.~Hill, Engineering clock transitions in molecular
  lanthanide complexes, Journal of the American Chemical Society 146~(16)
  (2024) 11083--11094.
\newblock \href {https://doi.org/10.1021/jacs.3c09353}
  {\path{doi:10.1021/jacs.3c09353}}.

\bibitem{Ghosh2021-Mn3dimer}
T.~Ghosh, J.~Marbey, W.~Wernsdorfer, S.~Hill, K.~A. Abboud, G.~Christou,
  Exchange-biased quantum tunnelling of magnetization in a [mn3]2 dimer of
  single-molecule magnets with rare ferromagnetic inter-mn3 coupling, Physical
  Chemistry Chemical Physics 23~(14) (2021) 8854--8867.
\newblock \href {https://doi.org/10.1039/d0cp06611g}
  {\path{doi:10.1039/d0cp06611g}}.

\bibitem{Gakiya-Teruya2025-Hopzdo4}
M.~Gakiya-Teruya, R.~Stewart, L.~Peng, S.~Liu, C.~Li, H.-P. Cheng, G.~K.-L.
  Chan, S.~Hill, M.~Shatruk, A 54.6 ghz clock transition in ho3+ electron spin
  qubits assembled into a metal–organic framework, Journal of the American
  Chemical Society 147~(27) (2025) 24068--24076.
\newblock \href {https://doi.org/10.1021/jacs.5c07796}
  {\path{doi:10.1021/jacs.5c07796}}.

\bibitem{Fursina2023-device}
A.~A. Fursina, A.~Sinitskii, Toward molecular spin qubit devices: Integration
  of magnetic molecules into solid-state devices, ACS Applied Electronic
  Materials 5~(7) (2023) 3531--3545.
\newblock \href {https://doi.org/10.1021/acsaelm.3c00472}
  {\path{doi:10.1021/acsaelm.3c00472}}.

\bibitem{Chiesa2024-device}
A.~Chiesa, P.~Santini, E.~Garlatti, F.~Luis, S.~Carretta, Molecular
  nanomagnets: a viable path toward quantum information processing?, Reports on
  Progress in Physics 87~(3) (2024).
\newblock \href {https://doi.org/10.1088/1361-6633/ad1f81}
  {\path{doi:10.1088/1361-6633/ad1f81}}.

\bibitem{Kragskow2023SpinPhonon}
J.~G.~C. Kragskow, A.~Mattioni, J.~K. Staab, D.~Reta, J.~M. Skelton, N.~F.
  Chilton, Spin--phonon coupling and magnetic relaxation in single-molecule
  magnets, Chem. Soc. Rev. 52 (2023) 4567--4585.
\newblock \href {https://doi.org/10.1039/D2CS00705C}
  {\path{doi:10.1039/D2CS00705C}}.

\bibitem{Lunghi2022ExactPredictions}
A.~Lunghi, S.~Sanvito, Toward exact predictions of spin-phonon relaxation
  times: An ab initio implementation of open quantum systems theory, Sci. Adv.
  8 (2022) eabn7880.
\newblock \href {https://doi.org/10.1126/sciadv.abn7880}
  {\path{doi:10.1126/sciadv.abn7880}}.

\bibitem{lunghi2026unified}
A.~Lunghi, A unified ab initio theory of spin-phonon relaxation and decoherence
  uncovers fast dephasing in magnetic molecules, Science Advances 12~(12)
  (2026) eaeb3868.
\newblock \href {https://doi.org/10.1126/sciadv.aeb3868}
  {\path{doi:10.1126/sciadv.aeb3868}}.

\bibitem{Saito2000QME}
K.~Saito, S.~Takesue, S.~Miyashita, Energy transport in the integrable system
  in contact with various types of phonon reservoirs, Physical Review E 61~(3)
  (2000) 2397--2409.
\newblock \href {https://doi.org/10.1103/PhysRevE.61.2397}
  {\path{doi:10.1103/PhysRevE.61.2397}}.

\bibitem{Nakano2001Fe6}
H.~Nakano, S.~Miyashita, Magnetization process of nanoscale iron cluster,
  Journal of the Physical Society of Japan 70~(7) (2001) 2151--2157.
\newblock \href {https://doi.org/10.1143/jpsj.70.2151}
  {\path{doi:10.1143/jpsj.70.2151}}.

\bibitem{Breuer2002OpenQS}
H.-P. Breuer, F.~Petruccione, The Theory of Open Quantum Systems, Oxford
  University Press, Oxford, 2002.

\bibitem{Rousochatzakis2005PulsedField}
I.~Rousochatzakis, M.~Luban, Master equations for pulsed magnetic fields:
  Application to magnetic molecules, Physical Review B 72 (2005) 134424.
\newblock \href {https://doi.org/10.1103/PhysRevB.72.134424}
  {\path{doi:10.1103/PhysRevB.72.134424}}.

\bibitem{Chilton2013PHI}
N.~F. Chilton, R.~P. Anderson, L.~D. Turner, A.~Soncini, K.~S. Murray, {PHI}: A
  powerful new program for the analysis of anisotropic monomeric and
  exchange-coupled polynuclear $d$- and $f$-block complexes, J. Comput. Chem.
  34 (2013) 1164--1175.
\newblock \href {https://doi.org/10.1002/jcc.23234}
  {\path{doi:10.1002/jcc.23234}}.

\bibitem{Stoll2006EasySpin}
S.~Stoll, A.~Schweiger, Easyspin, a comprehensive software package for spectral
  simulation and analysis in epr, Journal of Magnetic Resonance 178~(1) (2006)
  42--55.
\newblock \href {https://doi.org/10.1016/j.jmr.2005.08.013}
  {\path{doi:10.1016/j.jmr.2005.08.013}}.

\bibitem{qdmag_code}
S.~Liu, \href{https://github.com/shuanglongliu/qdmag}{\textit{qdmag}} (2026).
\newblock \href {https://doi.org/10.5281/zenodo.18850033}
  {\path{doi:10.5281/zenodo.18850033}}.
\newline\urlprefix\url{https://github.com/shuanglongliu/qdmag}

\bibitem{GutierrezFinol2023}
G.~M. Guti\'errez-Finol, S.~Gim\'enez-Santamarina, Z.~Hu, L.~E. Rosaleny,
  S.~Cardona-Serra, A.~Gaita-Ari\~no, Lanthanide molecular nanomagnets as
  probabilistic bits, npj Computational Materials 9 (2023) 196.
\newblock \href {https://doi.org/10.1038/s41524-023-01149-7}
  {\path{doi:10.1038/s41524-023-01149-7}}.

\bibitem{Hogben2011}
H.~J. Hogben, M.~Krzystyniak, G.~T.~P. Charnock, P.~J. Hore, I.~Kuprov, Spinach
  -- a software library for simulation of spin dynamics in large spin systems,
  Journal of Magnetic Resonance 208~(2) (2011) 179--194.
\newblock \href {https://doi.org/10.1016/j.jmr.2010.11.008}
  {\path{doi:10.1016/j.jmr.2010.11.008}}.

\bibitem{Johansson2013QuTiP}
J.~R. Johansson, P.~D. Nation, F.~Nori, {QuTiP} 2: A {Python} framework for the
  dynamics of open quantum systems, Comput. Phys. Commun. 184 (2013)
  1234--1240.
\newblock \href {https://doi.org/10.1016/j.cpc.2012.11.019}
  {\path{doi:10.1016/j.cpc.2012.11.019}}.

\bibitem{Spinach2026PulsedField}
I.~Kuprov, Spinach: spin-phonon dissipator, pulsed-field experiment, and
  pulsed-field magnetometry case studies,
  \url{https://github.com/IlyaKuprov/Spinach}, commit cc7bfb78, directory
  examples/giant\_spin/case\_studies, accessed 29 October 2026 (2026).

\bibitem{Strang1968}
G.~Strang, On the construction and comparison of difference schemes, SIAM
  Journal on Numerical Analysis 5~(3) (1968) 506--517.
\newblock \href {https://doi.org/10.1137/0705041} {\path{doi:10.1137/0705041}}.

\bibitem{Foldi2007}
P.~F\"oldi, M.~G. Benedict, J.~M. Pereira, F.~M. Peeters, Dynamics of molecular
  nanomagnets in time-dependent external magnetic fields: Beyond the
  {L}andau--{Z}ener--{S}t\"uckelberg model, Physical Review B 75 (2007) 104430.
\newblock \href {https://doi.org/10.1103/PhysRevB.75.104430}
  {\path{doi:10.1103/PhysRevB.75.104430}}.

\bibitem{Rudowicz2001SpinH}
C.~Rudowicz, S.~K. Misra, Spin-hamiltonian formalisms in electron magnetic
  resonance (emr) and related spectroscopies, Applied Spectroscopy Reviews
  36~(1) (2001) 11--63.
\newblock \href {https://doi.org/10.1081/asr-100103089}
  {\path{doi:10.1081/asr-100103089}}.

\bibitem{Hutchings1964PointCharge}
M.~Hutchings, Point-charge calculations of energy levels of magnetic ions in
  crystalline electric fields, in: F.~Seitz, D.~Turnbull (Eds.), Solid State
  Physics, Vol.~16, Academic Press, 1964, pp. 227--273.
\newblock \href {https://doi.org/10.1016/S0081-1947(08)60517-2}
  {\path{doi:10.1016/S0081-1947(08)60517-2}}.

\bibitem{Rudowicz2004ESO}
C.~Rudowicz, C.~Y. Chung, The generalization of the extended stevens operators
  to higher ranks and spins, and a systematic review of the tables of the
  tensor operators and their matrix elements, Journal of Physics: Condensed
  Matter 16~(32) (2004) 5825--5847.
\newblock \href {https://doi.org/10.1088/0953-8984/16/32/018}
  {\path{doi:10.1088/0953-8984/16/32/018}}.

\bibitem{Chibotaru2012SINGLEANI}
L.~F. Chibotaru, L.~Ungur, Ab initio calculation of anisotropic magnetic
  properties of complexes. i. unique definition of pseudospin hamiltonians and
  their derivation, The Journal of Chemical Physics 137~(6) (2012).
\newblock \href {https://doi.org/10.1063/1.4739763}
  {\path{doi:10.1063/1.4739763}}.

\bibitem{Grabert1988QBrown}
H.~Grabert, P.~Schramm, G.-L. Ingold, Quantum brownian motion: The functional
  integral approach, Physics Reports 168~(3) (1988) 115--207.
\newblock \href {https://doi.org/10.1016/0370-1573(88)90023-3}
  {\path{doi:10.1016/0370-1573(88)90023-3}}.

\bibitem{Campaioli2024QME}
F.~Campaioli, J.~H. Cole, H.~Hapuarachchi, Quantum master equations: Tips and
  tricks for quantum optics, quantum computing, and beyond, PRX Quantum 5~(2)
  (2024).
\newblock \href {https://doi.org/10.1103/PRXQuantum.5.020202}
  {\path{doi:10.1103/PRXQuantum.5.020202}}.

\bibitem{Blanes2009Magnus}
S.~Blanes, F.~Casas, J.~A. Oteo, J.~Ros, The magnus expansion and some of its
  applications, Physics Reports 470~(5-6) (2009) 151--238.
\newblock \href {https://doi.org/10.1016/j.physrep.2008.11.001}
  {\path{doi:10.1016/j.physrep.2008.11.001}}.

\bibitem{2020SciPy-NMeth}
P.~Virtanen, R.~Gommers, T.~E. Oliphant, M.~Haberland, T.~Reddy, D.~Cournapeau,
  E.~Burovski, P.~Peterson, W.~Weckesser, J.~Bright, S.~J. {van der Walt},
  M.~Brett, J.~Wilson, K.~J. Millman, N.~Mayorov, A.~R.~J. Nelson, E.~Jones,
  R.~Kern, E.~Larson, C.~J. Carey, {\.I}.~Polat, Y.~Feng, E.~W. Moore,
  J.~{VanderPlas}, D.~Laxalde, J.~Perktold, R.~Cimrman, I.~Henriksen, E.~A.
  Quintero, C.~R. Harris, A.~M. Archibald, A.~H. Ribeiro, F.~Pedregosa, P.~{van
  Mulbregt}, {SciPy 1.0 Contributors}, {{SciPy} 1.0: Fundamental Algorithms for
  Scientific Computing in Python}, Nature Methods 17 (2020) 261--272.
\newblock \href {https://doi.org/10.1038/s41592-019-0686-2}
  {\path{doi:10.1038/s41592-019-0686-2}}.

\bibitem{LeeSteppyMn3}
M.~Lee, P.~Rosa, H.~Tsai, W.~Nie, D.~Lubert-Perquel, S.~Hill, H.-P. Cheng,
  V.~Zapf, Quantum dynamics of an antiferromagnetic molecule reached in pulsed
  magnetic field, in preparation (2026).

\bibitem{Blum2012}
K.~Blum, Density matrix theory and applications, third edition. Edition,
  Springer series on atomic, optical, and plasma physics, 64, Springer, 2012.

\bibitem{Higham2005Pade}
N.~J. Higham, The scaling and squaring method for the matrix exponential
  revisited, SIAM Journal on Matrix Analysis and Applications 26~(4) (2005)
  1179--1193.
\newblock \href {https://doi.org/10.1137/04061101X}
  {\path{doi:10.1137/04061101X}}.

\bibitem{AlMohy2009Pade}
A.~H. Al-Mohy, N.~J. Higham, A new scaling and squaring algorithm for the
  matrix exponential, SIAM Journal on Matrix Analysis and Applications 31~(3)
  (2009) 970--989.
\newblock \href {https://doi.org/10.1137/09074721X}
  {\path{doi:10.1137/09074721X}}.

\bibitem{AlMohy2011Action}
A.~H. Al-Mohy, N.~J. Higham, Computing the action of the matrix exponential,
  with an application to exponential integrators, SIAM Journal on Scientific
  Computing 33~(2) (2011) 488--511.
\newblock \href {https://doi.org/10.1137/100788860}
  {\path{doi:10.1137/100788860}}.

\bibitem{Saad1992Krylov}
Y.~Saad, Analysis of some krylov subspace approximations to the matrix
  exponential operator, SIAM Journal on Numerical Analysis 29~(1) (1992)
  209--228.
\newblock \href {https://doi.org/10.1137/0729014} {\path{doi:10.1137/0729014}}.

\bibitem{Sidje1998Expokit}
R.~B. Sidje, Expokit: A software package for computing matrix exponentials, ACM
  Transactions on Mathematical Software 24~(1) (1998) 130--156.
\newblock \href {https://doi.org/10.1145/285861.285868}
  {\path{doi:10.1145/285861.285868}}.

\end{thebibliography}

\end{document}